\documentclass[twocolumn,showpacs,preprintnumbers,amsmath,amssymb,floatfix]{revtex4}
\usepackage{epsfig}
\usepackage{wasysym}
\usepackage{amssymb}
\usepackage{slashed}
\usepackage{physics}
\usepackage{color}
\usepackage{booktabs}
\usepackage{array}
\usepackage{tabularx}
\usepackage{multirow}
\usepackage{longtable}
\usepackage{ragged2e}
\usepackage{braket,booktabs}
\usepackage{braket,amsmath}
\usepackage{xcolor}
\usepackage{graphicx}
\begin{document}

\newcommand{\bvec}[1]{\mbox{\boldmath ${#1}$}}
\title{Slowly Rotating and Tidal Deformation of Nonlocal Modified Tolman VII Star}
\author{Byon N. Jayawiguna}
\email[]{nugrahabyon312@gmail.com}
\author{Piyabut Burikham}
\email[]{piyabut@gmail.com}

\affiliation{High Energy Physics Theory Group, Department of Physics, Faculty of Science, Chulalongkorn University, Bangkok 10330, Thailand}

\date{\today}
\begin{abstract}
We investigate the moment of inertia, quadrupole deformation, and tidal deformation within the framework of nonlocal gravity, utilizing the exact modified Tolman-VII (NEMTVII) density model with an isotropic perfect fluid. The Love number~$(k_{2})$ is derived using standard even-parity perturbation theory. Additionally, we explore the observational implications by analyzing the tidal deformability parameter~$( \lambda_{\textrm{tid}} )$ in comparison with the constraints from GW170817, GW190425, PSR J0348+0432, and PSR J0740+6620. We found that the results are consistent with the tidal constraint when $\alpha \gtrsim 1.6$ with the small $\beta$. For slowly rotating object, the dimensionless moment of inertia~$( \bar{I} )$, rotational Love parameter~$( \bar{\lambda}_{\textrm{rot}} )$, and quadrupole moment~$( \bar{Q} )$ are fully determined by the perturbed metric. Our findings reveal that the nonlocal parameter~$( \beta )$ significantly affects the star radius. For a fixed $\beta$ and varying $\alpha$, the $I$-Love-$Q$ relations are found to be universal. For varying $\beta$, the $I$-Love-$Q$ relations become non-universal. 
\end{abstract}

\maketitle
\section{Introduction} \label{SectI}

The study of astrophysical objects such as neutron stars (NS) remains a compelling research due to the limited understanding of their physical properties in the high density regime~\cite{Ozel:2016oaf,Oertel:2016bki,Baym:2017whm,Lattimer:2019eez}. One approach is to combine theoretical models with observational data to constrain the Mass-Radius (M-R) relation and the Equation of State (EoS) of the star. Thus, the physical properties such as the moment of inertia, and the deformability characterized by the Love number and the quadrupole moment \cite{Love,Postnikov:2010yn} can be determined. Several efforts have been devoted to observing the mass and radius by Neutron Star Interior Composition Explorer (NICER) and providing the constraints on the realistic equation of state (EoS) \cite{Miller:2019cac,Riley:2019yda}. Observations of gravitational waves from binary neutron star mergers using second-generation ground-based detectors, e.g. Advanced LIGO, Advanced Virgo, and KAGRA might enable the measurement of the tidal Love number~\cite{Flanagan:2007ix,Hinderer:2007mb,Hinderer:2009ca,Damour:2012yf,LIGOScientific:2017vwq,LIGOScientific:2018hze} and other multipole moments.

In exploring tidal deformation from both Newtonian and General Relativity~(GR) perspectives, distinct methodologies emerge. In the Newtonian framework, tidal deformation arises when the companion mass exerts the external tidal field to another object in the neighborhood body zone area. This is quantitatively described by the total potential of the deformed body, which is consist of the potential of the unperturbed configuration combined with the contributions from the quadrupole moment and the external tidal field. However, in GR, tidal deformation describes the distortion of a massive body due to the gravitational influence of a nearby companion to the spacetime. Here, the deformation is governed by the curvature of spacetime, with the gravitational potential embedded in the $ g_{tt} $ component of the metric tensor. The theoretical setup to obtain the tidal aspect, related to the tidal Love number, was first developed by Flagan and Hinderer \cite{Flanagan:2007ix,Hinderer:2007mb}. Mathematically, the relativistic treatment of tidal deformation is expressed in the limit of large distances $r,$~\cite{Thorne:1997kt}
\begin{eqnarray}
\label{expansion}
-\frac{(1+g_{tt})}{2}&=&-\frac{M}{r}-\frac{3\mathcal{Q}_{ij}}{2r^3}\left(n^{i}n^{j}-\frac{1}{3}\delta^{ij}\right)+\mathcal{O}\left(\frac{1}{r^4}\right) \nonumber \\ &&+ \frac{1}{2}\mathcal{E}_{ij}x^{i}x^{j}+\mathcal{O}(r^3),
\end{eqnarray}
where $ n^{i}=x^{i}/r $. The first term in the right-hand side of eqn.~\eqref{expansion} represents the well-known monopole field of the star with mass $M$. The second (third) term represents the quadrupole (higher quadrupole) term respectively. The fourth term is the external tidal contribution.

On the other hand, when considering a rotating star, the interplay between rotation and gravitational effects becomes significant. For such stars, the field equations can be approximated by perturbing the non-rotating solution. In the context of multipole moments, the moment of inertia, quadrupole moment and the Love number are important parameters in explaining various aspects of the EoS within the star. The study of slowly rotating stars within the GR framework represents a well-established research supported by an extensive literature. Significant contributions to this literature include the formulation of the Hartle-Thorne metric \cite{Hartle:1967he,Hartle:1968si}, which provides a theoretical framework for describing the spacetime around a slowly rotating star. The latter work focused on extending the Schwarzschild solution to include the effects of slow rotation. Yagi and Yunes \cite{Yagi:2013awa,Yagi:2013bca} found that the moment of inertia, Love number, and the quadrupole deformation (``$I$-Love-$Q$") for realistic EoS satisfy a universal relation which are almost independent of the interior of the star.

To analyze the observational results, we need to obtain the interior solution of the star. There exists an extensive literature discussing the solution within the analytical framework. The simplest one is the constant density star (CDS) \cite{Schwarzschild,misner,Stuchlik:2000gey}. There are also the power-law polytropic fluid stars studied in e.g. Ref.~\cite{tooper,Stuchlik:2016xiq} and realistic EoS's in more specific multiquark models in e.g. Ref.~\cite{Burikham:2010sw,Burikham:2021xpn,Pinkanjanarod:2020mgi,Pinkanjanarod:2021qto,Demircik:2022uol}. While the assumption of constant energy density is a simplification, the Tolman VII model~\cite{Tolman:1939jz,Durgapal} is an analytic density model constructed in such a way that the density is positive at the center and drops to zero at the surface. It provides a baseline for understanding the structure and properties of more realistic objects comparing to the CDS model.  The stability of the model using the adiabatic index is discussed in Ref.~\cite{Moustakidis:2016ndw}. The dimensionless tidal, and the moment of inertia $I$ in the TVII model are studied in \cite{Lattimer:2000nx,Jiang:2020uvb}. It is reported that the profile of the tidal (when plotted against compactness) in this model is similar to the polytropic Equation of State (EoS): has a positive nonzero value at the low compactness and  decreases monotonically until the maximum compactness \cite{Postnikov:2010yn}.

Another literature discussing the TVII model is reported in \cite{Jiang:2019vmf} where an extra term is introduced to make the density model more realistic: Modified Tolman VII (MTVII). The modified density model is characterized by the free parameter $ \alpha $ and reduces to TVII when $ \alpha=1. $ The relevant range of the parameter is $ \alpha \in[0.4,~1.4] $ and $ \mathcal{C}\in[0.05,~0.35] $~\cite{Jiang:2020uvb} in order to cover wider range of NS profiles. With this physical range, analytic solution of $I$-Love-$C$ relation and the tidal deformability are explored~\cite{Jiang:2020uvb}. The post-Minkowskian recursive perturbation method~\cite{Chan:2014tva} is used to obtain the moment of inertia $I$ and the angular momentum per mass $\omega$. The dynamical stability under small and adiabatic radial perturbations of the MTVII model is extensively studied in Ref.~\cite{Posada:2021zxk}. They discovered that the new model remains stable against radial oscillations.  Nonetheless, incorporating a fully numerical solution is essential for comprehensive study of the EoS profile. Camilo \textit{et al.}~\cite{Posada:2022lij} derive exact numerical solution for the MTVII model, which we will call the exact modified Tolman VII (EMTVII) solution. They found that the physical range of the free parameter $\alpha$ can be extended to $\alpha\in[0,2]$.

In addition to GR, modified gravity theories present alternative descriptions of field equations. These theories, in general, incorporate additional fields or modify the gravitational action sector to accommodate deviations from GR. For example, the modified theory can be expressed with the scalar-tensor theory and $ f(\mathcal{R}) $ gravity \cite{Brans:1961sx,Damour:1996ke,Silva:2014ora,Manolidis:2014tis}. In these frameworks, the effects from the slowly rotating parameter and the tidal parameter may differ from GR predictions due to the effective Einstein field equations \cite{Sotani:2010dr,Sotani:2010re,Sotani:2012eb,Staykov:2014mwa,Silva:2014fca,Yazadjiev:2016pcb,Staykov:2018hhc,Murshid:2023xsw}. 

In the quest to unify quantum mechanics and GR, there is a large number of literature discussing various theoretical frameworks that challenge our understanding of spacetime and gravity. Among these approaches, nonlocal gravity, which is often associated with the Generalized Uncertainty Principle (GUP), is a possible candidate (see \cite{Battisti:2007jd,Shibusa:2007ju,Battisti:2007zg,Nozari:2009cs,Bojowald:2011jd,Huang:2012hia,Sprenger:2012uc,Isi:2013cxa,Ali:2013ma,Tawfik:2014zca,Bruneton:2016yws,Li:2020vqg,Jayawiguna:2022ftj} for review). It has been reported that the MTVII model can be applied to broader theory such as nonlocal gravity. In \cite{Jayawiguna:2023vvw}, the nonlocal extension of the MTVII model, referred to as NEMTVII, reveals that the incorporation of an additional nonlocal parameter leads to an increase in the stellar radius and hence decreases the compactness. Thus, there are fewer trapped quasinormal modes and gravitational waves, as the eigen-solution of the Regge-Wheeler (RW) equation possesses a shorter echo time. To complete the analyses, in this work we examine the slowly rotating approximation and the tidal deformation of the NEMTVII model \cite{Jayawiguna:2023vvw}. We also compare our results with the EMTVII model predicted by Camilo, \textit{et al.} \cite{Posada:2022lij} and discuss the observational constraints from from GW170817, GW190425, PSR J0348+0432, and the PSR J0740+6620. 

The structure of this paper is the following. In Section~\ref{Sec2}, we review the calculation of the moment of inertia, rotational Love number, and the quadrupole moment of a slowly rotating star. In Section~\ref{Sec3}, we consider the situation with $ \bar{\omega}=0 $. In Section~\ref{Sec4}, we implement the NEMTVII density model to the Einstein field equation. Plots of the background solution and the tidal Love number, $k_{2}$ and $\bar{\lambda}_{\textrm{tid}}$, the dimensionless multipole moments $\bar{I}$, $\bar{\lambda}_{\textrm{rot}}$, $Q$ are presented in Sect.~\ref{Sec5}. Section~\ref{Sec6} concludes our work.

\section{Slowly Rotating Star}     \label{Sec2}

\subsection{Zeroth-order}

In this section we provide a brief review of the basic zeroth-order solution, which will later be used as the background metric to construct slowly-rotating and tidally deformed stars. The starting point is the ansatz metric for non-rotating configuration which is described by the spherical coordinates
\begin{equation}
ds^2=-e^{\nu(r)}dt^2 + e^{\psi(r)}dr^2+r^2 d\theta^2+r^2\sin^2\theta d\phi^2,
\end{equation}
where $ e^{-\psi(r)}=1-2m(r)/r $. With this ansatz, the components of the Einstein field equation ($t,t $), $ (r,r) $, ($ r,r $)-($ \theta,\theta $) in zeroth-order are given by
\begin{eqnarray}
\label{masseq}
&& \frac{dm}{dr}= 4\pi r^2 \rho, \\ && \frac{d\nu}{dr} = \frac{8\pi r^3 p + 2m(r)}{r[r-2m(r)]}  \\  && \label{nueq} \left(\frac{e^{-\psi}-1}{r^2}\right)' + \left(\frac{e^{-\psi}\nu'}{2r}\right)' + e^{-(\psi+\nu)} \left(\frac{e^{\nu}\nu'}{2r}\right)'=0.\nonumber \\ 
\end{eqnarray}
The Tolman-Oppenheimer-Volkov (TOV) expression can be obtained from the radial part of the conservation of energy, $ \nabla_{\mu}T^{\mu r}=0, $
\begin{equation}
\label{tov}
\frac{dp}{dr}=-(\rho+p) \frac{m(r)+4\pi p r^3}{r[r-2m(r)]}.
\end{equation}
The field equations and TOV equation must be solved to determine the solution within the stellar interior. Outside the star, where the pressure and density approach zero, the solution is the Schwarzschild solution
\begin{equation}
e^{\nu(r)}=e^{-\psi(r)}=1-\frac{2M}{r},~~~\textrm{for $ r>R, $}
\end{equation}
where the total mass is defined by $ M = m(R). $ Both interior and exterior solutions of the star are matched at the surface $r=R$ of the star denoted by $\Sigma_{0}$. Thus we can express the variable and its derivative as
\begin{eqnarray}
\label{conditions}
[\psi]=0,~~~[\nu]=0,~~~[\nu']=0,
\end{eqnarray}
where $ [x] $ denotes the difference between the value of $ x $ inside and outside evaluated at $ \Sigma_{0}, $ i.e., $ [x]=x^{-}|_{\Sigma_{0}}-x^{+}|_{\Sigma_{0}}. $ The prime denotes a derivative with respect to radial coordinate $r.$ In this convention, we use $ +~(-) $ to describe quantities in the outside (inside) of the star.

\subsection{Hartle-Thorne Formalism}

We now adopt the Hartle-Thorne formalism~\cite{Hartle:1967he,Hartle:1968si} to describe a slowly rotating body. Initially, we decompose the relevant line element, expressed in spherical coordinates~(the relation to Boyer-Lindquist coordinates is given in Ref.~\cite{Hartle:1968si})
\begin{eqnarray}
\label{hartleansatz}
ds^2&=&-e^{\nu}[1+2h_{0}(r)+2h_{2}(r)P_{2}(\cos\theta)]dt^2 \nonumber \\ &&+e^{\psi}\bigg\lbrace 1+\frac{e^{\psi}}{r}[2m_{0}(r)+2m_{2}(r)P_{2}(\cos\theta)] \bigg\rbrace dr^2  \nonumber \\ && + r^2[1+2(\nu_{2}-h_{2})P_{2}(\cos\theta)] \nonumber \\ && \times[d\theta^2+\sin^2\theta(d\phi-\omega dt)^2],
\end{eqnarray}
where $ P_{2}(\cos\theta) $ is a Legendre polynomial and $\omega(r)$ represents the angular velocity of the local inertial frame relative to a distant observer. The latter variable is also related to the dragging effect. $ h_{0} $ and $ m_{0} $ represent the spherically symmetric deformation of the star $ (l=0) $, while $ h_{2}, $ $ m_{2}, $ and $ \nu_{2} $ represent the quadrupole deformation $ (l=2) $.  

In this work, we assume the matter configuration to be isotropic perfect fluid with the 4-velocity given by~\cite{Hartle:1967he,Hartle:1968si}
\begin{eqnarray}
u^{t}&=& (-g_{tt}-2\Omega g_{t\phi}-g_{\phi\phi}\Omega^2)^{-1/2},\\ \nonumber &=& e^{-\nu/2}\bigg[1+\frac{1}{2}r^2 \sin^2\theta(\Omega-\omega)^2 e^{-\nu} \\ \nonumber && -h_{0}-h_{2}P_{2}\bigg], \\ \nonumber 
u^{\phi} &=& \Omega u^{t}, \\ \nonumber 
u^{r} &=& 0,\\ \nonumber
u^{\theta} &=& 0,
\end{eqnarray}
where $\Omega$ is the angular speed constant. From the ansatz, it is convenient to introduce, $ \bar{\omega}=\Omega-\omega $. Subsequently, using the perturbed metric in \eqref{hartleansatz}, we will address each order in separate subsections.

\subsubsection{First-order}

We utilize axisymmetric perturbations to examine the star to the linear order in $\bar{\omega}$. At the first order in perturbation, the only non-vanishing component of the Einstein field equation is ($ t,\phi $), which can be written as 
\begin{eqnarray}
\frac{d}{dr}\left[r^4 j \frac{d\bar{\omega}}{dr}\right]+4r^3 \frac{dj}{dr}\bar{\omega}=0, 
\end{eqnarray}
where $ j\equiv e^{-(\psi+\nu)/2}. $ After some algebra, this can be rewritten as
\begin{eqnarray}
\label{omegaeq}
&&\frac{d^2\bar{\omega}}{dr^2}+4\left[ \frac{1-\pi r^2 (\rho+p)e^{\psi}}{r}  \right] \frac{d\bar{\omega}}{dr}\nonumber \\ && -16\pi(\rho+p)e^{\psi}\bar{\omega}=0
\end{eqnarray}
The equation above can be evaluated with two boundary conditions $ \bar{\omega}(0)=1 $ and $ (d\bar{\omega}/dr)|_{r=0}=0. $ On the other hand, outside the star where the density and pressure goes to zero ($ \rho=p=0 $), Eqn.\eqref{omegaeq} can be solved analytically to give 
\begin{equation}
\bar{\omega}^{+}=\Omega-\frac{2J}{r^3},~~~\textrm{with}~~~j(r)=1,
\end{equation}
where $ J $ is the integration constant related to the angular momentum of the star. The complete solution can be determined by applying matching condition between both sides of the solution. Thus, one must have $ [\bar{\omega}]=0 $ to obtain
\begin{eqnarray}
\Omega = \bar{\omega}(R)+\frac{2J}{R^3},
\end{eqnarray}
and $ [\bar{\omega}']=0 $ leads to
\begin{eqnarray}
J = \frac{1}{6} R^4 \left(\frac{d\bar{\omega}}{dr}\right)_{r=R}.
\end{eqnarray}
The moment of inertia can be obtained from the relation $ I\equiv J/\Omega. $ In presenting our results, we use dimensionless moment of inertia
\begin{equation}
\bar{I}\equiv \frac{I}{M^3}.
\end{equation}

\subsubsection{Second-order}

The next-to-leading order expansion of slowly rotating equations leads to the determination of quadrupole moment. We first present the sets of differential equations for both regions of interest and then discuss how to solve it based on Hartle approach \cite{Hartle:1967he,Hartle:1968si}. The second-order field equations can be written as 
\begin{eqnarray}
\label{nu2}
\frac{d\nu_{2}}{dr}&=&-\left(\frac{d\nu}{dr}\right)h_{2} + \bigg(\frac{1}{r}+\frac{1}{2}\frac{d\nu}{dr}\bigg)\nonumber \\ &&\times \bigg[ \frac{1}{6} r^4 j^2 \left(\frac{d\bar{\omega}}{dr}\right)^2 -\frac{1}{3}r^3 \bar{\omega}^2 \frac{dj^2}{dr}\bigg],
\end{eqnarray}
\begin{eqnarray}
\label{h2}
\frac{dh_{2}}{dr}&=&\bigg\lbrace -\frac{d\nu}{dr}+\frac{r}{r-2m}\left(\frac{d\nu}{dr}\right)^{-1} \bigg[ 8\pi(\rho+p)-\frac{4m}{r^3} \bigg]  \bigg\rbrace h_{2} \nonumber \\ && -\frac{4\nu_{2}}{r(r-2m)} \left(\frac{d\nu}{dr}\right)^{-1} \nonumber \\ && + \frac{1}{6} \bigg[ \frac{r}{2} \left(\frac{d\nu}{dr}\right)-\frac{1}{(r-2m)} \left(\frac{d\nu}{dr}\right)^{-1}   \bigg] r^3 j^2 \left( \frac{d\bar{\omega}}{dr}  \right)^2 \nonumber \\ && - \frac{1}{3} \bigg[ \frac{r}{2} \left(\frac{d\nu}{dr}\right)+\frac{1}{(r-2m)} \left(\frac{d\nu}{dr}\right)^{-1}   \bigg] r^2 \frac{dj^2}{dr} \bar{\omega}^2.
\end{eqnarray}
Outside the star, Eqn.~\eqref{nu2} and \eqref{h2} can be integrated analytically 
\begin{eqnarray}
h_{2}^{+}=J^2 \left(\frac{1}{M r^3}+\frac{1}{r^4}\right) +  K Q_{2}^{2}\left(\frac{r}{M}-1\right),\nonumber \\
\label{outsidesol} \\
\nu_{2}^{+} = - \frac{J^2}{r^4}+ K \frac{2M}{\sqrt{r(r-2M)}} Q_{2}^{1}\left(\frac{r}{M}-1\right) \nonumber
\end{eqnarray}
where $ K $ and $ Q_{n}^{m} $ are the integration constant and the associated Legendre functions of the second kind, respectively.

However, by combining the two sets of equations presented in \eqref{nu2} and \eqref{h2} with the exterior solution given in \eqref{outsidesol}, one can numerically evaluate the solution and apply the regularity conditions. In practice, we employ Hartle's method \cite{Hartle:1967he,Hartle:1968si,Yagi:2013awa,Yagi:2013bca,Burikham:2021xpn} for solving these equations. The procedure is as follows: first decompose the interior solution into the sum of a homogeneous component, which involves arbitrary constants, and a particular component
\begin{equation}
\label{interior}
h_{2} = A h_{2}^{h} + h_{2}^{p},~~~\textrm{and}~~~\nu_{2}=A\nu_{2}^{h}+\nu_{2}^{p},
\end{equation}
where the particular part is given by Eqn.~\eqref{nu2} and \eqref{h2} with the following near-origin conditions
\begin{eqnarray}
h_{2}^{p}&=& ar^2,\\
\nu_{2}^{p}&=& \frac{2\pi}{3} \left[(\rho_{c}+p_{c})j_{c}^2-(\rho_{c}+3p_{c})a\right]r^2
\end{eqnarray}
where $ a $ is a constant. On the other hand, the homogeneous part can be obtained from
\begin{equation}
\frac{d\nu_{2}^{h}}{dr} =- \left(\frac{d\nu}{dr}\right)h_{2}^{h},
\end{equation}
\begin{eqnarray}
\frac{dh_{2}^{h}}{dr}&=& \bigg\lbrace -\frac{d\nu}{dr}+\frac{r}{r-2m}\left(\frac{d\nu}{dr}\right)^{-1} \bigg[ 8\pi(\rho+p)-\frac{4m}{r^3} \bigg]  \bigg\rbrace h_{2}^{h} \nonumber \\ && -\frac{4\nu_{2}^{h}}{r(r-2m)} \left(\frac{d\nu}{dr}\right)^{-1},
\end{eqnarray}
with the near-origin conditions
\begin{equation}
h_{2}^{h}=Br^2,~~~\textrm{and}~~~\nu_{2}^{h} = -\frac{2\pi}{3}\left(\rho_{c}+3p_{c}\right) B r^4,
\end{equation}
where $ B $ is a constant. For a given value of $ a $ and $ B, $ both homogeneous and particular part can be solved numerically. Hence, the unknown constant $ K $ and $ A $ are determined by the algebraic equations 
\begin{equation}
[h_{2}]=0,~~~\textrm{and}~~~[\nu_{2}]=0,
\end{equation}
where the ($ h_{2},\nu_{2} $) are given by Eqn.~\eqref{outsidesol} and \eqref{interior}. The dimensionless quadrupole moment can be calculated from 
\begin{eqnarray}
\bar{Q}\equiv -\frac{Q^{(\textrm{rot})} M}{(I\Omega)^2}= 1+\frac{8K}{5} \left(\frac{M^2}{I\Omega}\right)^2,
\end{eqnarray}
which depends on the constant $K$ determined from the matching conditions. When $ K\rightarrow 0 $, the corresponding value of $ \bar{Q} $ approaches BH limit value.

At the second order, we can also define dimensionless rotational Love number in terms of moment of inertia and quadrupole moment
\begin{equation}
\bar{\lambda}^{(\textrm{rot})}=\bar{I}^2 \bar{Q}.
\end{equation}
Thus, $ \bar{\lambda}^{(\textrm{rot})} $ is determined by the first-order quantity $\bar{I}$ and the second-order perturbation in $\bar{Q}$.

\section{Tidal Love Number and Tidal Deformability}  \label{Sec3}

In this section, we consider a non-spinning, spherically symmetric object immersed in an external tidal field $ \mathcal{E}_{ij}$. The tidal effect enters at the second order. Due to its influence, the star will deform and develop a multipolar structure in response to the tidal field. Thus we can set $ \bar{\omega}=0 $ in the second-order field equation \eqref{nu2} and \eqref{h2}. After eliminating $ \nu_{2} $, we can obtain the master equation for $ h_{2} $ \cite{Hinderer:2007mb,Hinderer:2009ca}:
\begin{eqnarray}
\label{Hint}
&&h_{2}''(r) + \bigg\lbrace \frac{2}{r} + e^{\psi}\left[\frac{2m(r)}{r^2}+4\pi r (p-\rho)\right]\bigg\rbrace h_{2}'(r)\nonumber\\&& +\left[4\pi e^{\psi} \left( 5\rho+9p+\frac{\rho+p}{\gamma}   \right)-\frac{6e^{\psi}}{r^2} - \nu'^2    \right] h_{2}(r)=0.\nonumber \\
\end{eqnarray}
The prime denotes the derivative with respect to $ r $, and $\gamma\equiv dp/d\rho$. The function $ h_{2}(r) $ hence represents the solution for the interior of the star. With the quantity $ y\equiv rh_{2}'(r)/h_{2}(r), $ we can also transform the master equation (\ref{Hint}) into
\begin{eqnarray}
\label{Hinty}
\frac{dy(r)}{dr}&=&\frac{y(r)}{r}-\frac{y(r)^2}{r}\nonumber \\ && -\bigg\lbrace \frac{2}{r} + e^{\psi}\left[\frac{2m(r)}{r^2}+4\pi r (p-\rho)\right]\bigg\rbrace  y(r)\nonumber \\ && + \left[\frac{6e^{\psi}}{r^2}-4\pi e^{\psi} \left( 5\rho+9p+\frac{\rho+p}{\gamma}   \right) + \nu'^2    \right] r. \nonumber \\
\end{eqnarray}
Since this equation is first-order differential equation, we only need one boundary condition near the center and match the interior solution with the exterior solution. Outside the star ($ T_{\mu\nu}=0 $), eqn~\eqref{Hint} reduces to
\begin{equation}
\label{Hext}
h_{2}''(r)+\left(\frac{2}{r}-\psi'\right)h_{2}'(r)-\left(\frac{6e^{\psi}}{r^2}+\psi'^2\right)h_{2}(r)=0,
\end{equation} 
and the solution is a linear combination of Legendre functions $ c_{1}Q_{2}^2(r/M-1) $ and $ c_{2}P_{2}^{2}(r/M-1) $. We can obtain the coefficients $ c_{1} $ and $ c_{2} $ by employing the asymptotic behaviour and matching with the star's local asymptotic rest frame at large $ r $ in \eqref{expansion}. The results are
\begin{equation}
c_{1}=\frac{15\lambda_{\textrm{tid}} \mathcal{E}_{m}}{8M^3},~~~\textrm{and}~~~c_{2}=\frac{M^2\mathcal{E}_{m}}{3},
\end{equation}
where $\mathcal{E}_{m}$ is the component of the tidal field $ \mathcal{E}_{ij} $ in \eqref{expansion} which can be decomposed into series of spherical harmonic basis (with $ l=2 $) as
\begin{equation}
\mathcal{E}_{ij} = \sum^{2}_{m=-2} \mathcal{E}_{m}\mathcal{Y}_{ij}^{2m},
\end{equation}
and $\lambda_{\textrm{tid}}$ is the tidal deformability parameter defined by the linear response~\cite{Hinderer:2007mb,Hinderer:2009ca}
\begin{equation}
\label{qande}
\mathcal{Q}_{m}=-\lambda_{\textrm{tid}}\mathcal{E}_{m},
\end{equation}
when the multipole moments $\mathcal{Q}_{ij}$ is also decomposed into
\begin{equation}
\mathcal{Q}_{ij} = \sum^{2}_{m=-2} \mathcal{Q}_{m}\mathcal{Y}_{ij}^{2m}.
\end{equation}
Note that the spherical harmonic basis $ \mathcal{Y}_{ij}^{2m} $ is given by
\begin{equation}
Y^{m}_{2}(\theta,\phi)=\mathcal{Y}^{2m}_{ij}n^{i}n^{j}.
\end{equation}
The $ Y^{m}_{2}(\theta,\phi) $ is the spherical harmonic and $ n\equiv(\sin\theta\cos\phi,~\sin\theta\sin\phi,~\cos\theta)$. 
The tidal deformability is related to the dimensionless tidal Love number $k_{2}$ by
\begin{equation}
k_{2}=\frac{3}{2}\lambda_{\textrm{tid}} R^{-5},
\end{equation}
and the dimensionless tidal parameter can be defined as
\begin{equation}
\label{dimleslambda}
\bar{\lambda}_{\textrm{tid}}=\lambda_{\textrm{tid}}/M^5= \frac{2}{3} k_{2}\mathcal{C}^{-5},
\end{equation}
where $\mathcal{C}=M/R$ is the compactness. In terms of $y_{R}=y(r=R)$ the tidal Love number $ k_{2} $ is 
\begin{equation}
\label{k2}
k_{2} =-\frac{4\mathcal{C}^5}{15}\left( \frac{R\frac{dP_{2}^{2}}{dr}\big|_{r=R}  -y_{R} P_{2}^{2}}{R\frac{dQ_{2}^{2}}{dr}\big|_{r=R}-y_{R} Q_{2}^{2}}\right)=\frac{B_{1}}{B_{2}},
\end{equation}
where
\begin{equation}
B_{1}= \frac{8\mathcal{C}^5}{5} (1-2\mathcal{C})^2[2+2\mathcal{C}(y_{R}-1)-y_{R}],
\end{equation}
and 
\begin{eqnarray}
B_{2}&=& 2\mathcal{C}[6-3y_{R}+3\mathcal{C}(5y_{R}-8)]+4\mathcal{C}^3[13-11y\nonumber \\ && +\mathcal{C}(3y_{R}-2)+2\mathcal{C}^2(1+y_{R})] +3(1-2\mathcal{C})^2[2-y_{R}\nonumber \\ &&+2\mathcal{C}(y_{R}-1)]\times\log|1-2\mathcal{C}|.\nonumber \\ 
\end{eqnarray}
Thus, the tidal Love number $ k_{2} $ is obtained by evaluating Eqn.~\eqref{Hinty} with the boundary condition $ y(r\approx0)=2 $ and substituting $ y_{R}$ to the right-hand side of Eqn.~\eqref{k2}.

\section{The Mass Model}  \label{Sec4}

In previous sections, we consider the slowly rotating approximation and the tidal deformation of the star and analyze the moment of inertia, quadrupole moment, and tidal deformation parameters. We will now shift our focus to the mass~(density) model used to describe the (EoS) of the star. For this purpose, we consider an isotropic perfect fluid as the matter configuration in the original modified Tolman VII (MTVII) density model, characterized by the central density $\rho_{c}$, the parameter $\alpha$, and the radius $R$. Subsequently, the nonlocal effect will be incorporated into the mass model.

\subsection{Modified Tolman VII (EMTVII)}

Jiang and Yagi \cite{Jiang:2019vmf} (see also \cite{Posada:2022lij}) propose a modified Tolman VII (MTVII) model
\begin{equation}
\label{rhomod}
\rho_{\rm mod}(r) = \rho_{c} \left[ 1-\alpha \frac{r^2}{R^2} + (\alpha-1) \frac{r^4}{R^4}  \right]
\end{equation}
with a new parameter $ \alpha $. When $ \alpha=1, $ the density model is reduced to the Tolman VII model. The central density for modified model, $ \rho_{c}, $ can be written as
\begin{equation}
\rho_{c} = \frac{105 \mathcal{C}}{8\pi R^2(10-3\alpha)},
\end{equation}
and the mass and metric $g_{rr}$ read
\begin{eqnarray}
m_{\rm mod}(r) &=& \frac{\mathcal{C} R}{2 (10-3\alpha)} \left[35\frac{r^3}{R^3} - 21\alpha\frac{r^5}{R^5} + 15 (\alpha-1) \frac{r^7}{R^7}   \right],\nonumber \\ \label{lambda}
e^{-\psi_{\rm mod}} &=& 1- \frac{2m_{\textrm{mod}}(r)}{r},
\end{eqnarray}
where ``mod" refers to the modified model corresponding to the density profile described in \eqref{rhomod}. According to Ref.~\cite{Jiang:2019vmf}, the metric $ \nu(r) $ is no longer analytic since the expression is more complicated compared with the TVII model. Camilo Posada, \textit{et al.} \cite{Posada:2022lij} developed the numerical solution of the MTVII model, which they called Exact Modified Tolman VII (EMTVII). The existence of compact objects in EMTVII shows wider physically allowed range,~$ \alpha \in [0,2] $. Moreover, they showed that the maximum compactness based on the finiteness of pressure, $ \mathcal{C}_{\textrm{max}}, $ is achieved when $ \alpha=0 $, $ \mathcal{C}_{\textrm{max}}=0.4 $. This value is higher than TVII model $ (\alpha=1,~\mathcal{C}_{\textrm{max}}=0.386) $.

\subsection{Nonlocal Gravity Coupled with Modified Tolman VII (NEMTVII)}

We have examined several neutron star models within the framework of general relativity~(GR), which remains successful across a wide range of energy scales. However, unresolved questions remain. The existence of curvature singularity inside the black hole and naked singularity in the GR framework indicate that GR is incomplete. Canonical quantum gravity could possibly regulate the singularity by quantum fluctuations at the Planck scale. However, when gravity is canonically quantized, the calculations lead to ultraviolet~(UV) divergences that cannot be renormalized \cite{tHooft:1974toh,Deser:1974cz,Deser:1974cy,Goroff:1985th}. Nonlocal theories, such as those inspired by the UV complete quantum gravity theory~(see e.g. Ref.~\cite{Gaete:2010sp,Mureika:2010je,Moffat:2010bh,Modesto:2010uh,Nicolini:2012eu}) offer a promising avenue toward quantum gravity. Nonlocal gravity modifies the Einstein-Hilbert action by introducing d’Alembertian operator that allows curvature to have nonlocal interaction. Nonlocal gravity uses the entire function of the operator to regulate gravity at very short distances, avoiding infinities and unphysical behavior while still agreeing with Einstein’s gravity at large distances \cite{Moffat:2010bh}. It is thus interesting to couple the density model, EMTVII, with this kind of nonlocal gravity, so that we can consider nonlocal gravity effects on the compact object as explored in Ref.~\cite{Jayawiguna:2023vvw}. 

To summarize, the nonlocality is encoded in the action
\begin{equation}
    S=\frac{1}{16\pi} \int d^{4}x \sqrt{-g} \mathcal{A}^{2}(\Box) R + S_{\textrm{matter}},
\end{equation}
where $R$ is a Ricci scalar and the $\mathcal{A}^{2}$ is an operator, often chosen to be an entire function \cite{Isi:2013cxa}. The nonlocal Einstein field equation can be written in the following form \cite{Nicolini:2012eu,Gaete:2010sp,Mureika:2010je,Moffat:2010bh,Modesto:2010uh}
\begin{eqnarray}
\label{efeng2}
R_{\mu\nu}-\frac{1}{2}g_{\mu\nu}R = 8\pi  \mathcal{T}_{\mu\nu},
\end{eqnarray}
where $ \mathcal{T}_{\mu\nu}\equiv \mathcal{A}^{-2}(\Box)T_{\mu\nu} $. Here, the operator $ \Box_{x}= \beta g_{\mu\nu}\nabla^{\mu}\nabla^{\nu}, $ is dimensionless D'Alembertian operator and $\beta$ is the nonlocal parameter. Note that in this work, since we use the geometric unit with distance measured in km, the unit of $\beta$ is km$^2$. Eqn.\eqref{efeng2} denotes the ordinary gravity coupled to the generalized matter with nonlocal effects. The energy-momentum tensor is modified to
\begin{equation}
\label{EMT}
\mathcal{T}_{\mu\nu} = (\tilde{\rho}+\tilde{p})u_{\mu}u_{\nu} + \tilde{p} ~ g_{\mu\nu},
\end{equation}
and the component of the Einstein field equation is the same as Eqn.~\eqref{masseq}-\eqref{tov} with the tilde sign on the pressure and the density. Thus, the nonlocal density and the pressure will be coupled with the entire function $\mathcal{A}^{-2}(\Box)$, which can be determined by choosing the nonlocal model. However, choosing the entire function is essential for ensuring the consistency of the theory and physical validity. In particular, the entire function must be chosen to prevent the existence of the ghost, and effectively suppress high-energy (UV) effects without altering the behavior at low energies (IR)~\cite{Modesto:2011kw,Biswas:2011ar}.  To the best of our knowledge, there is no experimental information about quantum gravity hence there are no restrictions for choosing a suitable form of $\mathcal{A}^{-2}(\Box)$ \cite{Isi:2013cxa}. At the Planck scale, quantum gravity suggests a minimal length, though the standard uncertainty principle allows probing arbitrarily small scales. The Generalized Uncertainty Principle~(GUP) addresses this by introducing a fundamental minimal length, implying that spacetime becomes effectively nonlocal. In this work, we choose the entire function that supports the higher momenta based on the GUP model \cite{Isi:2013cxa,Kempf:1994su,Sprenger:2012uc} with 
\begin{equation}
\label{entirefunc}
\mathcal{A}(\Box)=\sqrt{1-\Box}.
\end{equation}
Moreover, the differential operator can be expressed further using the Schwinger representation~\cite{Isi:2013cxa}
\begin{equation}
    \hat{\Delta}^{\gamma} = \frac{1}{\Gamma(-\gamma)}\int_{0}^{\infty} ds~s^{-(\gamma+1) e^{-s\hat{\Delta}}},
\end{equation}
where $\hat{\Delta}$ is a generic differential operator. It means that the d'Alembertian operator is evaluated using an integral representation. Based on the GUP model, we have
\begin{equation}
    \mathcal{A}^{-2}(\Box)=(1-\Box)^{-1} = \int_{0}^{\infty} ds ~e^{-s(1-\Box)},
\end{equation}
where $\hat{\Delta}=1-\Box$, and $\gamma=-1$. Thus, the modified density can be written as \cite{Jayawiguna:2023vvw}
\begin{eqnarray}
\label{rhong}
\tilde{\rho}= \mathcal{A}^{-2} \rho &=& \label{rong} \frac{1}{2\sqrt{\beta}x} \int_{0}^{R} dx'~x'~\rho(x') \left[ e^{-\frac{|x-x'|}{\sqrt{\beta}}}- e^{-\frac{|x+x'|}{\sqrt{\beta}}}\right].\nonumber\\
\end{eqnarray}
Given the absolute value in the expression, it is necessary to decompose it into the interior and exterior regions and integrated with appropriate intervals. With the MTVII model presented in \eqref{rhomod}, gives
\begin{eqnarray}
\label{rhoint}
\tilde{\rho}_{\textrm{int}} (r) &=& \frac{\rho_{c}~e^{-(r+R)/\sqrt{\beta}}}{r/\sqrt{\beta}} \bigg\lbrace \left(1-e^{2r/\beta}\right) \bigg[ (\alpha-2)\nonumber \\ &+& (7\alpha-10)\frac{\sqrt{\beta}}{R} +3(9\alpha-10)\frac{\beta}{R^2} +60(\alpha-1)\frac{\beta^{3/2}}{R^3} \nonumber \\ &+&60(\alpha-1)\frac{\beta^2}{R^4} \bigg]+ \frac{e^{(r+R)/\sqrt{\beta}}~r}{\sqrt{\beta}}  \bigg[1-\alpha \frac{r^2}{R^2}\nonumber \\ &+&(\alpha-1) \frac{r^4}{R^4} -6\alpha \frac{\beta}{R^2}+120(\alpha-1) \frac{\beta^2}{R^4}\nonumber \\ &+&20(\alpha-1)\frac{\beta}{R^2}   \bigg]\bigg\rbrace,
\end{eqnarray}
and
\begin{eqnarray}
\label{rhoext}
\tilde{\rho}_{\textrm{ext}} (r)&=& \frac{\rho_{c}~ e^{-\frac{(r+R)}{\sqrt{\beta}}}}{r/\sqrt{\beta}} \bigg\lbrace \left(1-e^{2R/\sqrt{\beta}}\right)\bigg[(\alpha-2)\nonumber \\ && +3(9\alpha-10)\frac{\beta}{R^2} +60(\alpha-1)\frac{\beta^2}{R^4}  \bigg]\nonumber \\ &&+\left(1+e^{2R/\sqrt{\beta}}\right) \bigg[(7\alpha-10)\frac{\sqrt{\beta}}{R}\nonumber \\ &&+60 (\alpha-1)\frac{\beta^{3/2}}{R^3} \bigg]   \bigg\rbrace,
\end{eqnarray}
where $\tilde{\rho}_{\textrm{int}}$ and $\tilde{\rho}_{\textrm{ext}}$ are the interior and exterior nonlocal density. Important to note that the quantity $\tilde{\rho}_{\textrm{ext}}$ is not located outside of the star. Instead, it corresponds to the smearing surface. The modified nonlocal density profile is characterized by the star-radius parameter $ R, $~(due to nonlocal smearing effect, this is smaller than the actual radius of the star) nonlocal parameter $ \beta $ and the energy density value at the center. The mass can be written as 
\begin{eqnarray}
\label{massint}
m_{\textrm{int}} (r) &=& 4\pi\rho_{c}\bigg\lbrace \frac{(\alpha-1)}{7}\frac{r^7}{R^4} +\frac{r^5}{R^2}\left[\frac{4\beta}{R^2}(\alpha-1)-\frac{\alpha}{5}\right] \nonumber \\ && + \frac{r^3}{3}\left[1-\frac{6\alpha\beta}{R^2}+\frac{120 (\alpha-1)\beta^2}{R^4}\right]\nonumber\\&&-\left[e^{\frac{r-R}{\sqrt{\beta}}}(\beta r-\beta^{3/2}) + e^{-\frac{r+R}{\sqrt{\beta}}}(\beta r+\beta^{3/2}) \right]\nonumber \\ && \times\bigg[(\alpha-2)+(7\alpha-10)\frac{\sqrt{\beta}}{R} \nonumber \\ && +\frac{3\beta}{R^2}(9\alpha-10)+\frac{60\beta^{3/2}(\alpha-1)}{R^3}\nonumber \\ &&+\frac{60\beta^2(\alpha-1)}{R^4} \bigg] \bigg\rbrace,
\end{eqnarray}
and
\begin{eqnarray}
\label{massext}
m_{\textrm{ext}} (r)&=& 4\pi \rho_{c} \bigg[e^{-\frac{(r+R)}{\sqrt{\beta}}} (\beta r+\beta^{3/2}) - e^{-\frac{2R}{\sqrt{\beta}}} (\beta R + \beta^{3/2})\bigg]\nonumber \\ && \times\bigg\lbrace \left(e^{2R/\sqrt{\beta}}-1\right)\bigg[(\alpha-2) \nonumber \\ &&+3(9\alpha-10)\frac{\beta}{R^2} +60(\alpha-1)\frac{\beta^2}{R^4}  \bigg] \nonumber \\ && -\left(1+e^{2R/\sqrt{\beta}}\right) \bigg[(7\alpha-10)\frac{\sqrt{\beta}}{R}\nonumber \\ && +60 (\alpha-1)\frac{\beta^{3/2}}{R^3} \bigg] \bigg\rbrace.
\end{eqnarray}
Note that the exponential tail of the mass function $m_{\textrm{ext}}(r)$ in the exterior of the star signifies the smearing of the star surface. Numerical analysis of this model demonstrates that incorporating nonlocality parameter $ \beta $ leads to the existence of the extension region in both density and mass profile due to nonlocal smearing effect~\cite{Jayawiguna:2023vvw}. 
To obtain the actual star radius is to set our input into \eqref{rhoint}-\eqref{massext} and the radius ($ R_{N} $) can be determined when the density approaches very small cutoff value,~$ 10^{-7}~{\rm km}^{-2}$. The compactness is then given by $ \mathcal{C}=M_{N}/R_{N}. $ The second-order differential equation $ \nu(r) $ in \eqref{nueq} can be solved under the boundary condition in Eqn.~\eqref{conditions} with $ \Sigma_{0}=R_{N}$. The nonlocal pressure then can be determined using the Tolman-Oppenheimer-Volkoff (TOV) equation simultaneously. 

The nonlocal model allows the analysis of two distinct scenarios of compactness within the stellar profile. The first scenario corresponds to a compactness derived from a finite pressure condition. If the compactness exceeds the maximum allowable value ($ \mathcal{C}>\mathcal{C}_{\textrm{max}} $), both the pressure and the metric become divergent. The second scenario corresponds to the maximum compactness constrained by the causal limit (the subluminal speed of sound, $ \displaystyle{\frac{d\tilde{p}}{d\tilde{\rho}}<1} $). (see the details in Figure 1 and the corresponding Table in \cite{Jayawiguna:2023vvw}).

\section{Results and Discussions}   \label{Sec5}

In this section, we present numerical results for star profile, the Love number~($k_{2}$), dimensionless tidal deformability~($\bar{\lambda}_{\textrm{tid}}$, $\bar{\lambda}_{\text{rot}}$), moment of inertia~($\bar{I}$), and quadrupole deformation~($\bar{Q}$) using the NEMTVII model, and compare with the values from the EMTVII model. Plots of the $I$-Love-$Q$ relations are also presented.
\begin{figure}[t!]
	\centering
	\includegraphics[width=1\linewidth]{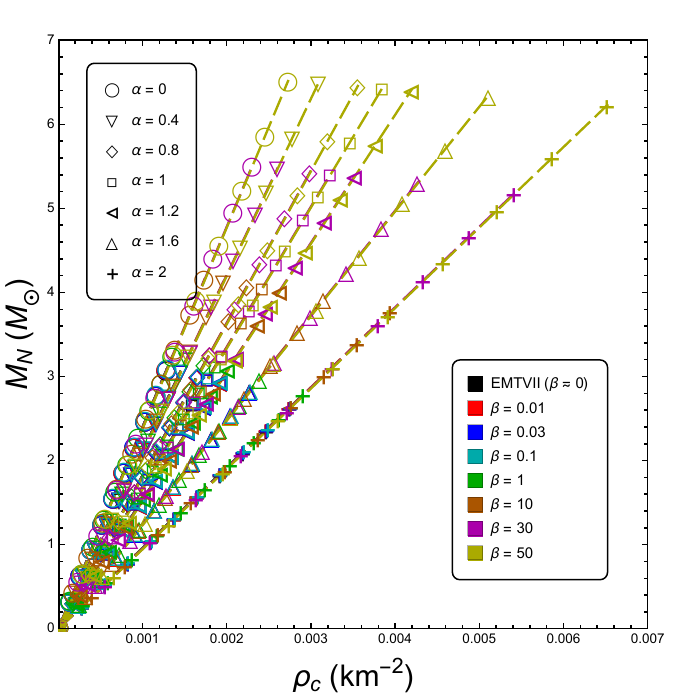}
	\includegraphics[width=1\linewidth]{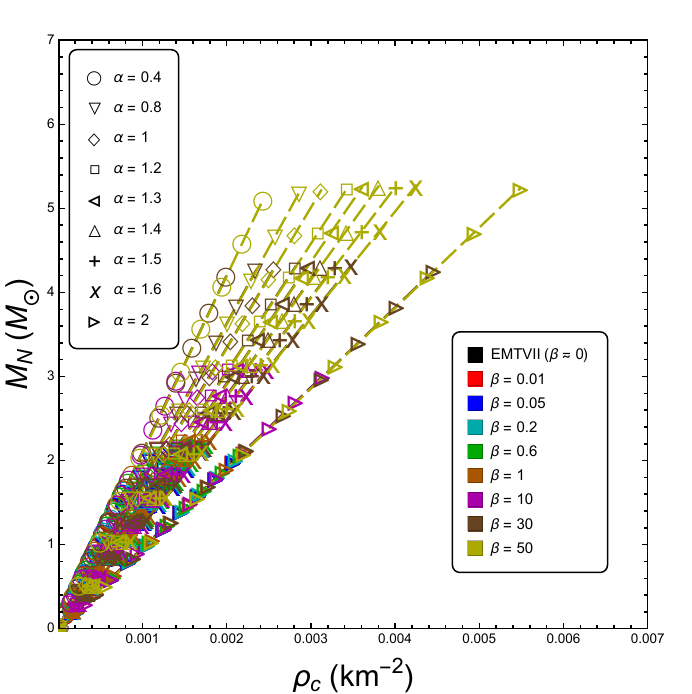}
	\caption{Mass vs central density based on the finite pressure criteria [Top] and within the subluminal speed of sound [Bottom]. The unit of $\beta$ is km$^2$.}
	\label{fig:1}
\end{figure}
We start with a thorough analysis of the unperturbed metric in NEMTVII model. The focus is on examining the mass profile (in solar masses) against the central density, as depicted in Figure~\ref{fig:1}. The one at the top (bottom) is the profile based on finiteness pressure (causal limit). For both cases, the central density varies from very small value until the maximum value. The highest central density is reached when $ \alpha=2. $ On the other hand, as the $ \beta $ increses, the mass becomes higher. This is due to the nonlocal effects in $ \tilde{\rho}_{\textrm{ext}} $ and $ \tilde{m}_{\textrm{ext}}, $ which increase the mass and radius of the nonlocal star.

\begin{figure}[t!]
	\centering
	\includegraphics[width=1\linewidth]{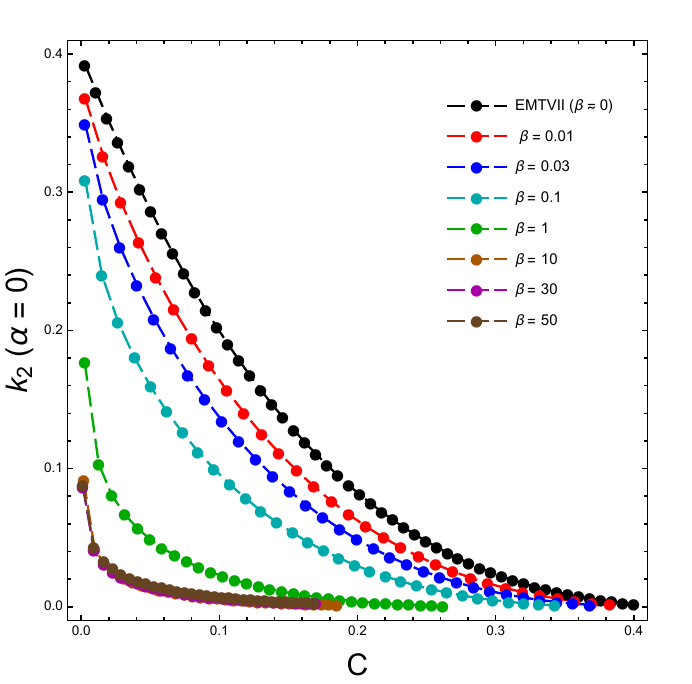}
	\includegraphics[width=1\linewidth]{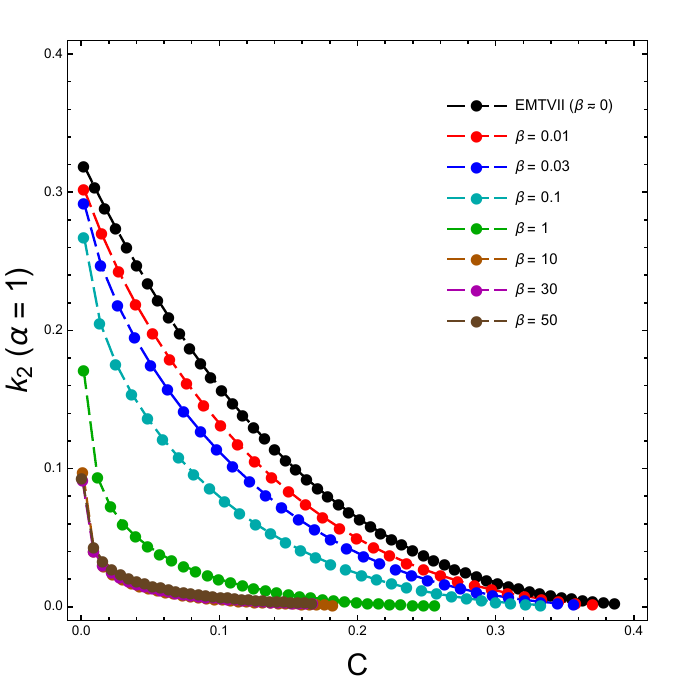}
	\caption{Tidal Love number, $ k_{2} $ vs $ \mathcal{C} $ for $\mathcal{C}\leq\mathcal{C}_{max}$ beyond which the pressure becomes divergent. Each colour represents the Love number for different value of $ \beta $ with black~(GR, $\beta=0$), red~($ \beta=0.01 $), blue~($ \beta=0.03 $), cyan~($ \beta=1 $), dark orange~($ \beta=10 $), magenta~($ \beta=30 $), and brown~($ \beta=50 $) line. }
	\label{fig:2}
\end{figure}
\begin{figure}[t!]
	\centering
	\includegraphics[width=1\linewidth]{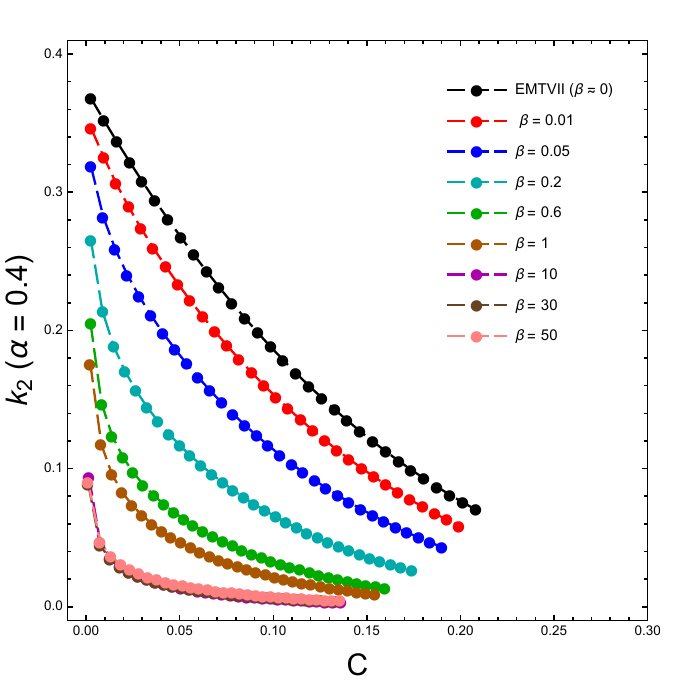}
	\includegraphics[width=1\linewidth]{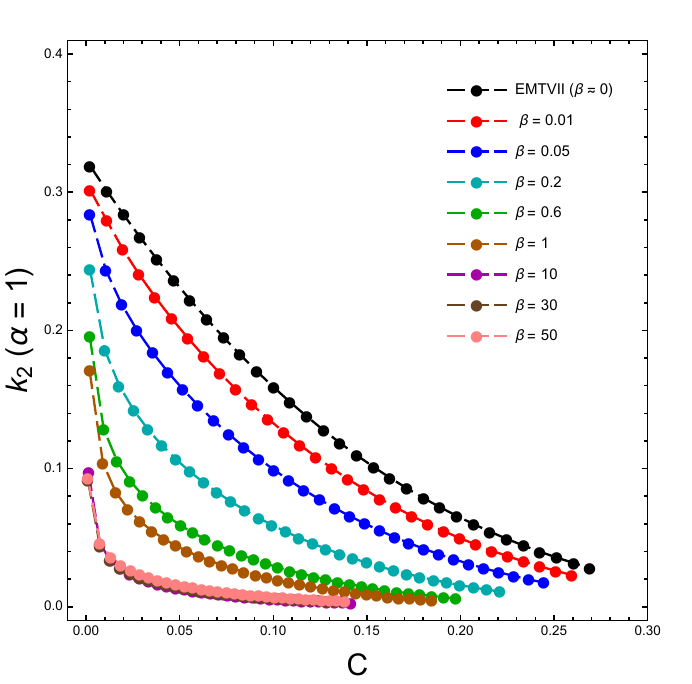}
	\caption{Tidal Love number, $ k_{2} $ vs $ \mathcal{C} $ for compactness within the causal limit. Each colour represents the Love number for different value of $ \beta $ with black~(GR, $\beta=0$), red~($ \beta=0.01 $), blue~($ \beta=0.05 $), cyan~($ \beta=0.2 $), dark orange~($ \beta=1 $), magenta~($ \beta=10 $), brown~($ \beta=30 $), and pink~($ \beta=50 $) line.}
	\label{fig:3}
\end{figure}
\begin{figure}[t!]
	\centering
	\includegraphics[width=1\linewidth]{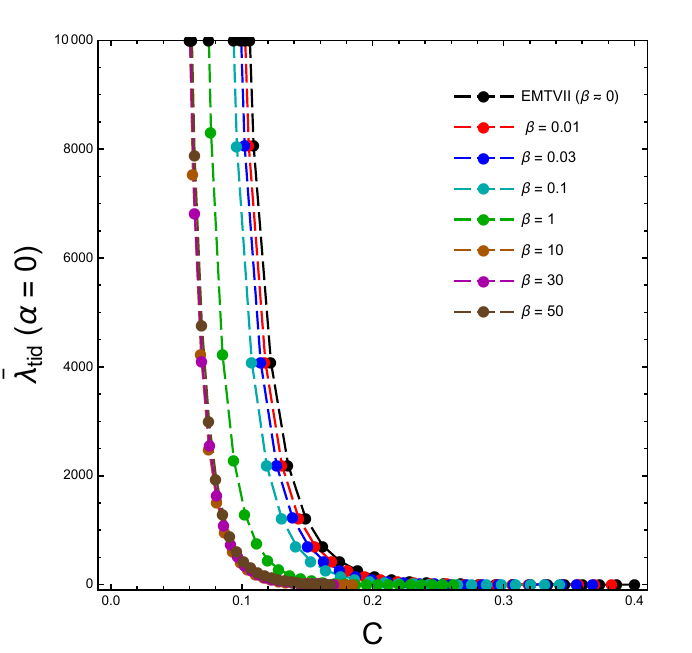}
	\includegraphics[width=1\linewidth]{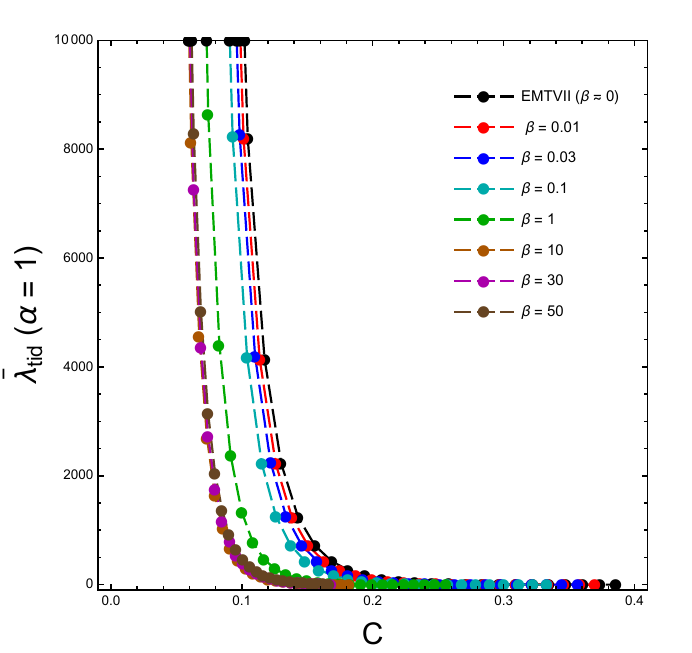}
	\caption{Dimensionless tidal deformability, $ \bar{\lambda}_{\textrm{tid}}, $ vs $ \mathcal{C} $ for $\mathcal{C}\leq\mathcal{C}_{max}$ beyond which the pressure becomes divergent in NEMTVII model. Each colour represents the dimensionless tidal deformability for different value of $ \beta $ with black~(GR, $\beta=0$), red~($ \beta=0.01 $), blue~($ \beta=0.03 $), cyan~($ \beta=1 $), dark orange~($ \beta=10 $), magenta~($ \beta=30 $), and brown~($ \beta=50 $) line.}
	\label{fig:4}
\end{figure}
\begin{figure}[t!]
	\centering
	\includegraphics[width=1\linewidth]{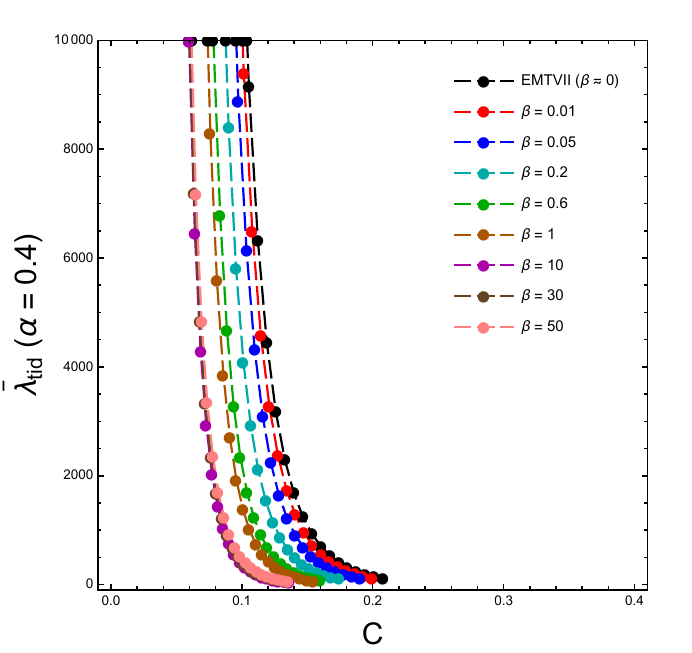}
	\includegraphics[width=1\linewidth]{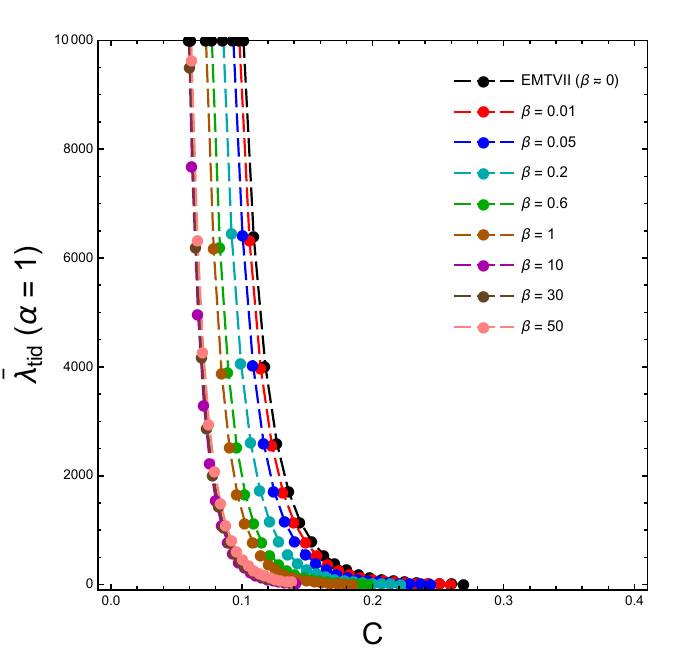}
	\caption{Dimensionless tidal deformability, $ \bar{\lambda}_{\textrm{tid}} $ vs $ \mathcal{C} $ for compactness within the causal limit. Each colour represents the Love number for different value of $ \beta $ with black~(GR, $\beta=0$), red~($ \beta=0.01 $), blue~($ \beta=0.05 $), cyan~($ \beta=0.2 $), dark orange~($ \beta=1 $), magenta~($ \beta=10 $), brown~($ \beta=30 $), and pink~($ \beta=50 $) line. }
	\label{fig:5}
\end{figure}

Figure~\ref{fig:2} (Figure~\ref{fig:3}) illustrates tidal Love number versus compactness for $ \mathcal{C}\leq\mathcal{C}_{\textrm{max}} $ (within the causal limit). The Love number $ k_{2} $ quantifies the deformation of a star in response to tidal forces. Specifically, a smaller Love number suggests that the star is more centrally condensed \cite{Hinderer:2009ca}. In NEMTVII model, $k_{2}$ is influenced by both free and nonlocal parameters. For $( \alpha = 1 )$, the profile represented by the black line corresponds to the TVII case. The plot shows that the nonlocal Love number decreases monotonically as the compactness approaches its maximum value, exhibiting similar behaviour to that observed in polytropic models (see Fig.~1 in Ref.~\cite{Hinderer:2009ca}). As we increase $\beta$, the profile of $ k_{2} $ decreases. It can be physically expected due to the higher central density as depicted in Figure~\ref{fig:1} when we increase the nonlocal parameter. However, we can also infer that the nonlocal Love number always has a non-negative value across the entire range of compactness. Similar behaviour appears in the Figure~\ref{fig:4} and Figure~\ref{fig:5} where we plot the relationship between $ \bar{\lambda}_{\textmd{tid}} $ and the compactness $ \mathcal{C}. $ As $ \beta $ increases, the profile shifts to the left, indicating that the larger $ \beta $ has lower $ \bar{\lambda}_{\textrm{tid}} $ at a given $ \mathcal{C} $. 
\begin{figure}
	\centering
	\includegraphics[width=0.84\linewidth]{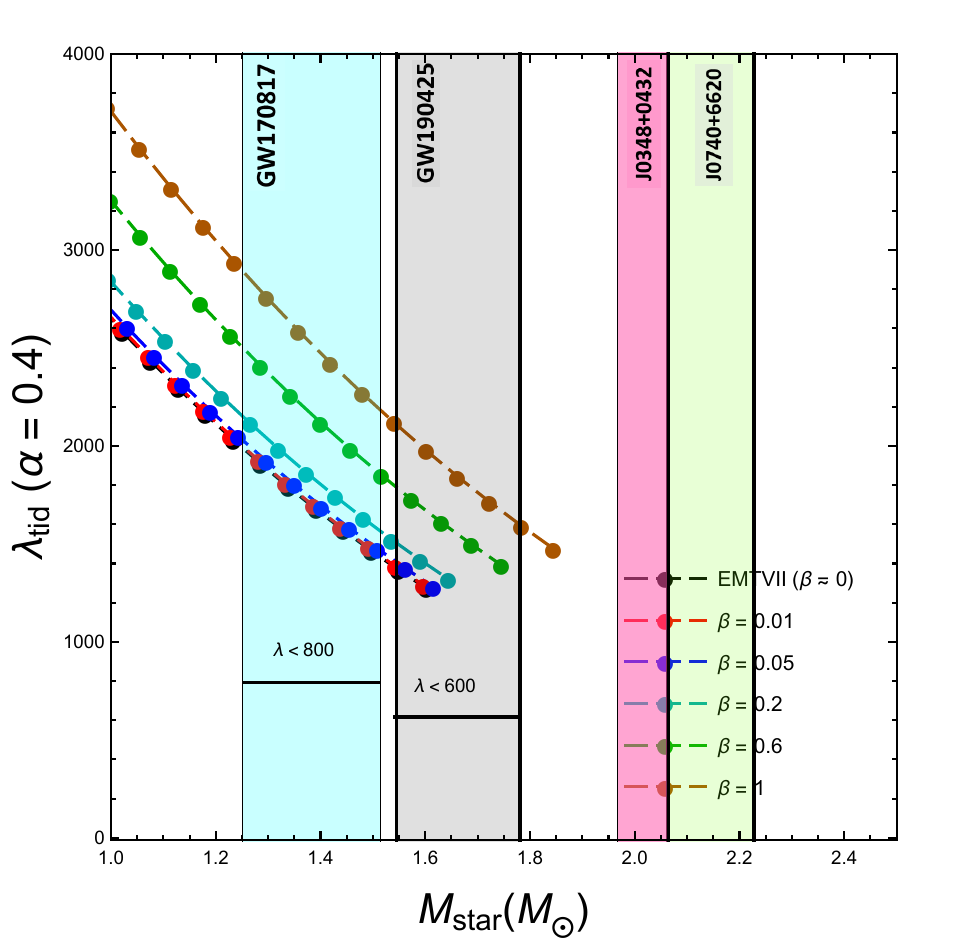}
	\includegraphics[width=0.84\linewidth]{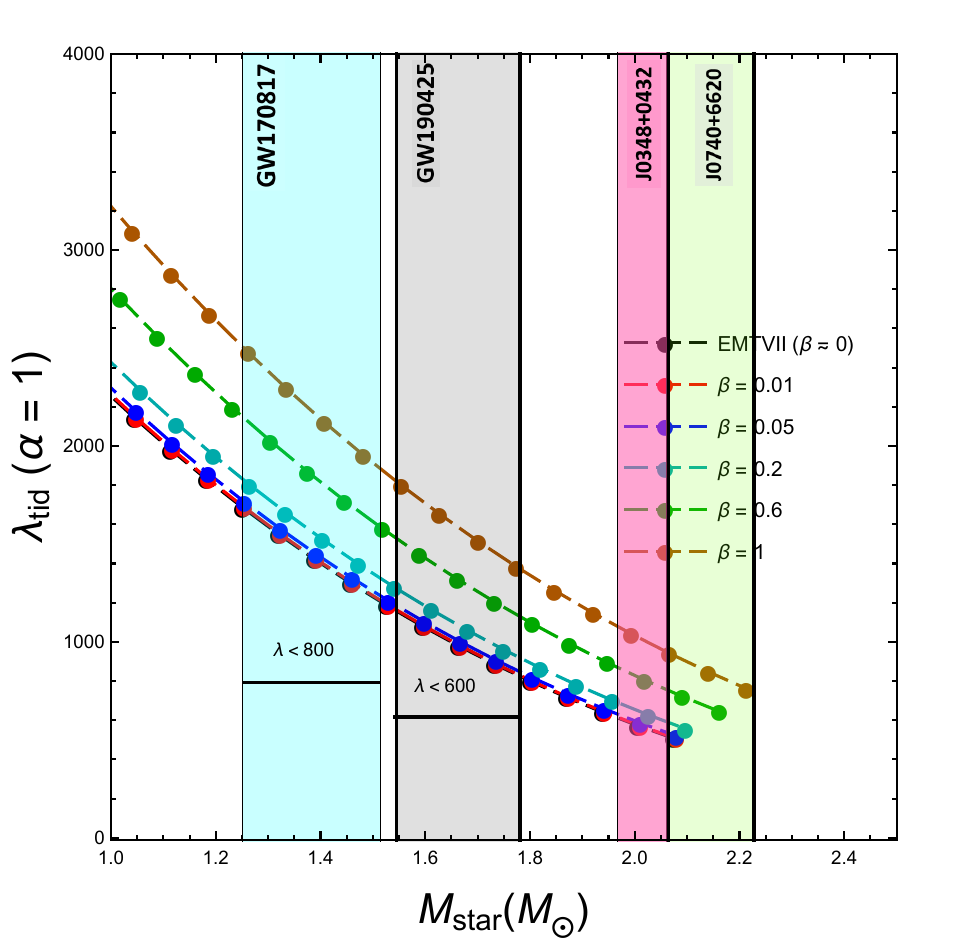}       
	\caption{The tidal deformability vs mass for $\alpha = 0.4, 1$ compared to the current observation in GW170817 \cite{LIGOScientific:2017vwq}, GW190425 \cite{LIGOScientific:2020aai}, PSR J0348+0432 \cite{Antoniadis:2013pzd}, and PSR J0740+6620 \cite{NANOGrav:2019jur,Fonseca:2021wxt}.}
	\label{fig:6}
\end{figure}
\begin{figure}
	\centering
	\includegraphics[width=0.84\linewidth]{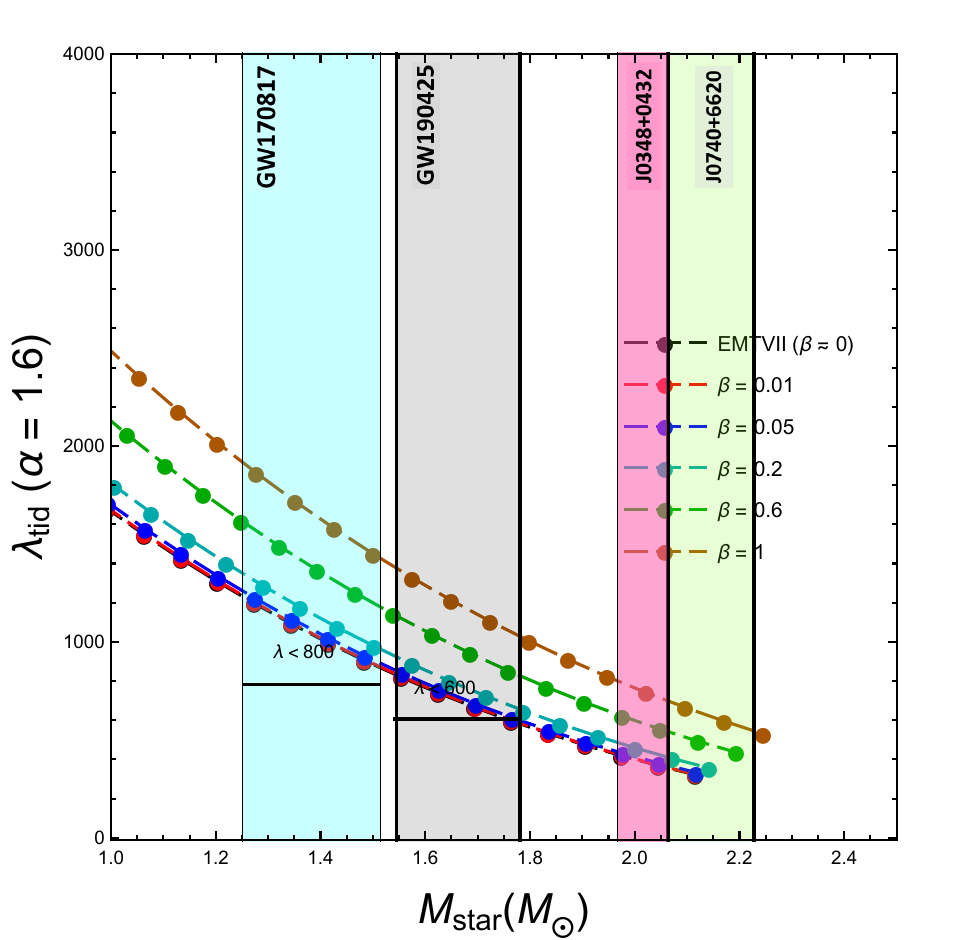}
	\includegraphics[width=0.84\linewidth]{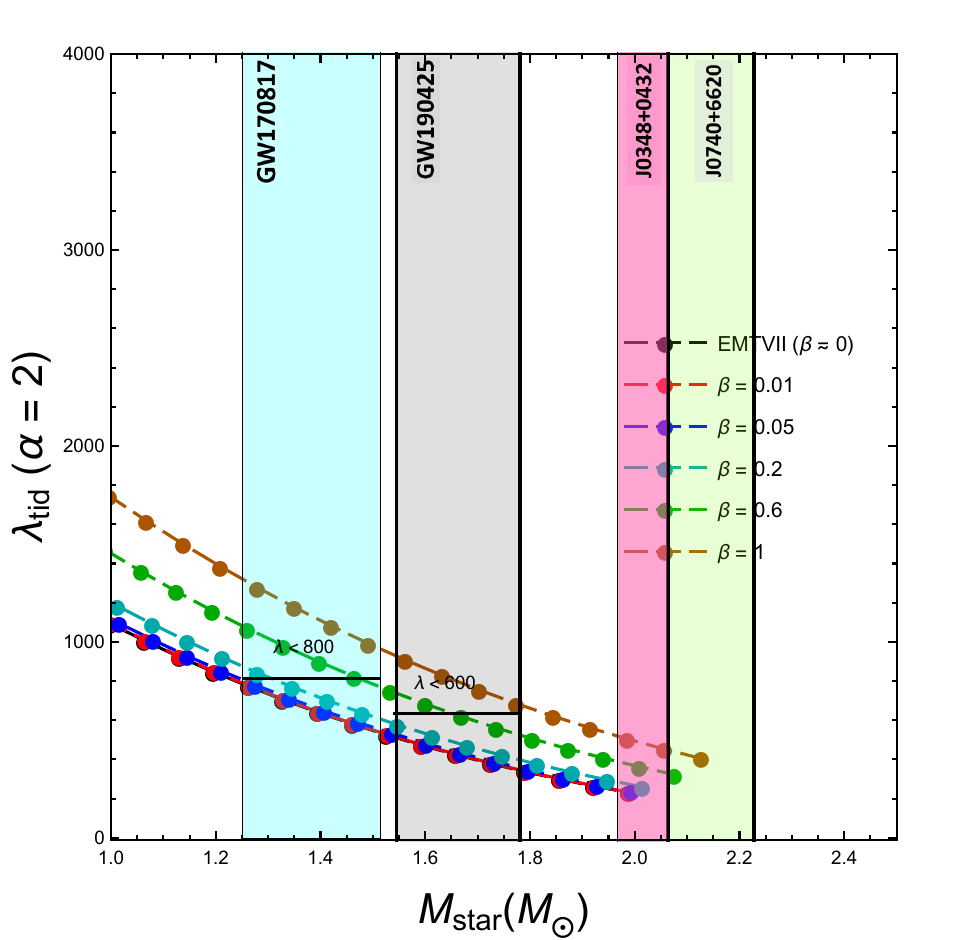}
	\caption{The tidal deformability vs mass for $\alpha = 1.6, 2$ compared to the current observation in GW170817 \cite{LIGOScientific:2017vwq}, GW190425 \cite{LIGOScientific:2020aai}, PSR J0348+0432 \cite{Antoniadis:2013pzd}, and PSR J0740+6620 \cite{NANOGrav:2019jur,Fonseca:2021wxt}.}
	\label{fig:7}
\end{figure}

The plots of tidal deformability $ \lambda_{\textrm{tid}} $ for the EMVTII and NEMTVII with respect to mass $ M $ are presented in Figure \ref{fig:6} and Figure \ref{fig:7}. We show star profiles only for the star within the causal limit with a black line for GR profile, red line for $ \beta=0.01, $ blue line for $ \beta=0.05, $ cyan line for $ \beta=0.2, $ green line for $ \beta=0.6 $, and the brown line for $ \beta=1$ km$^2$.
We also apply observational constraints from the gravitational wave event GW170817 \cite{LIGOScientific:2017vwq} and GW190425 \cite{LIGOScientific:2020aai} as shown in the Figure. The tidal deformability constraint from GW170817 (GW190425) is $ \lambda_{\textrm{tid}}<800 $ ($ \lambda_{\textrm{tid}}<600 $) \cite{LIGOScientific:2017vwq,LIGOScientific:2020aai} respectively. Additionally, we consider constraints from the observations: PSR J0348+0432 \cite{Antoniadis:2013pzd} and PSR J0740+6620 \cite{NANOGrav:2019jur,Fonseca:2021wxt}, where there are only mass constraints. For the profile with $ \alpha = 0.4 $, none of the values of $ \beta $ satisfy the observational constraints from PSR J0348+0432 and PSR J0740+6620. As $ \alpha $ increases, the tidal deformability profiles begin to intersect with the observational constraints. For $ \alpha = 1.6 $ and $ \alpha = 2 $, only a limited number of $ \beta $ values fall within the observational bounds set by the gravitational wave events GW170817 and GW190425. For a given $\alpha$, larger $\beta$ yields larger deformability, sufficiently large $\alpha \gtrsim 1.6$ and $\beta<1$ km$^2$ are required by observations.

\begin{figure}[t!]
	\centering
	\includegraphics[width=1\linewidth]{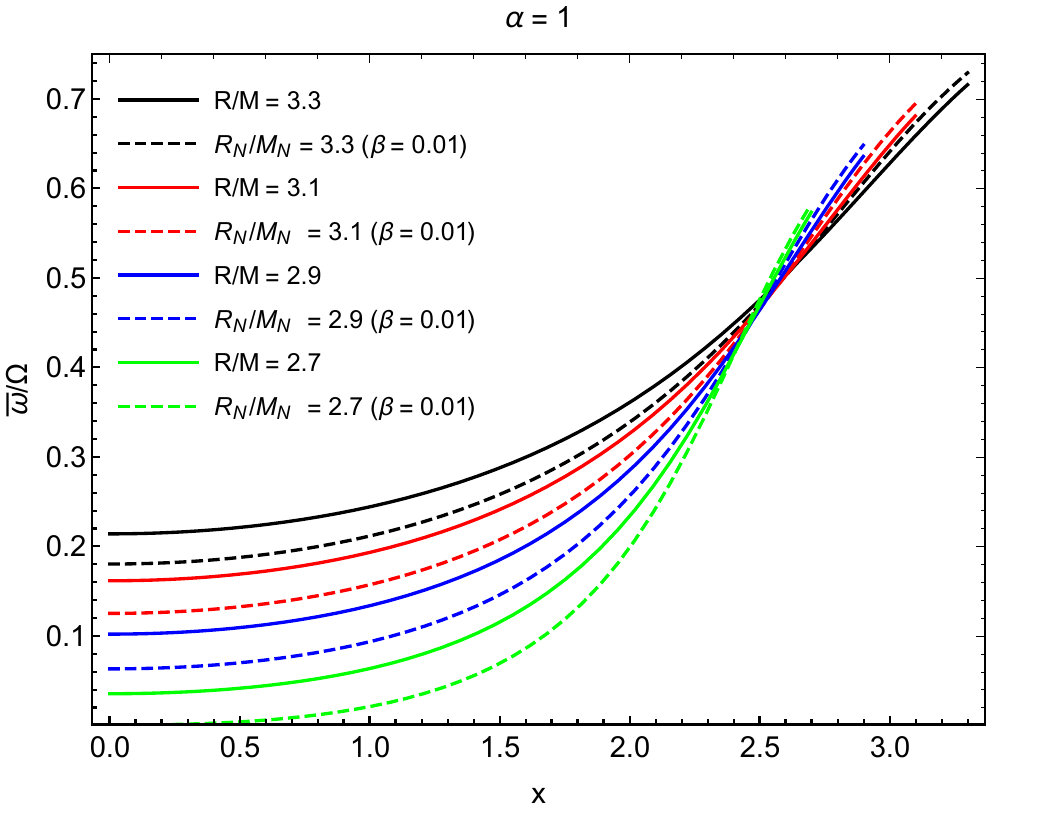}
    \includegraphics[width=1\linewidth]{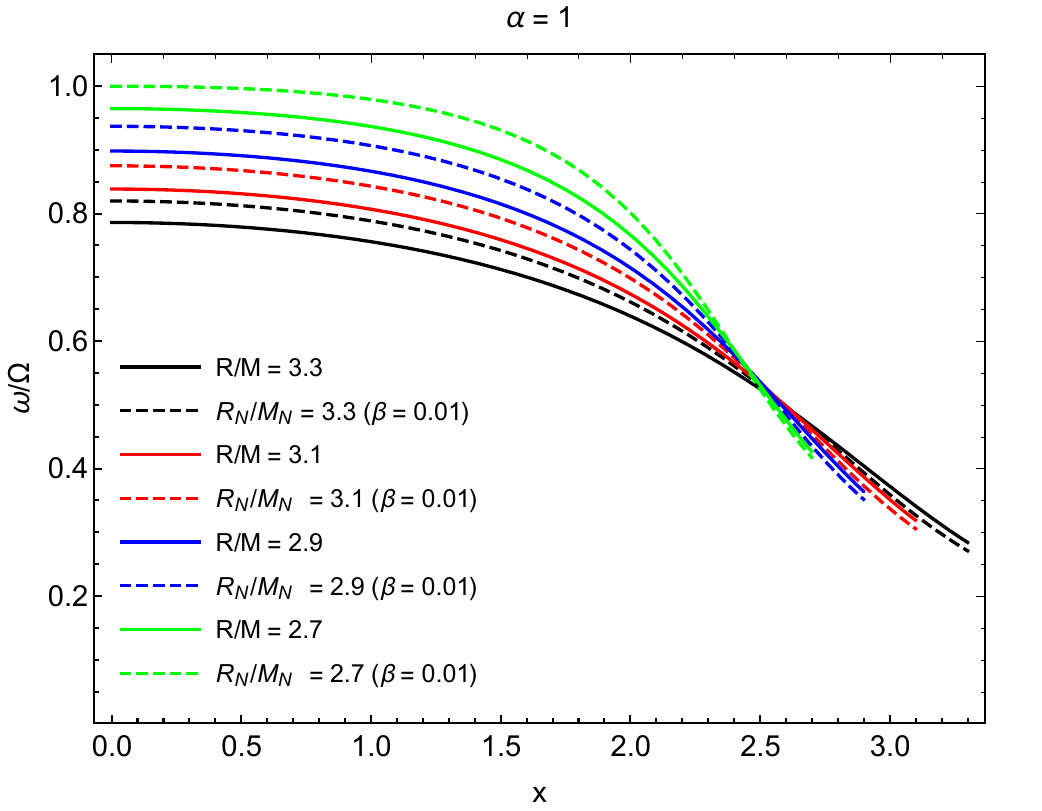}
	\caption{[Top] Angular velocity relative to the local inertial frame normalized by $\Omega $, $\bar{\omega}/\Omega$ and [Bottom] angular velocity normalized by $\Omega$, $\omega/\Omega$, in terms of the $x\equiv r/M$ ($x\equiv r/M_{N}$ for $\beta\neq0$).}
	\label{fig:omega}
\end{figure}
\begin{figure}[t!]
	\centering
	\includegraphics[width=1\linewidth]{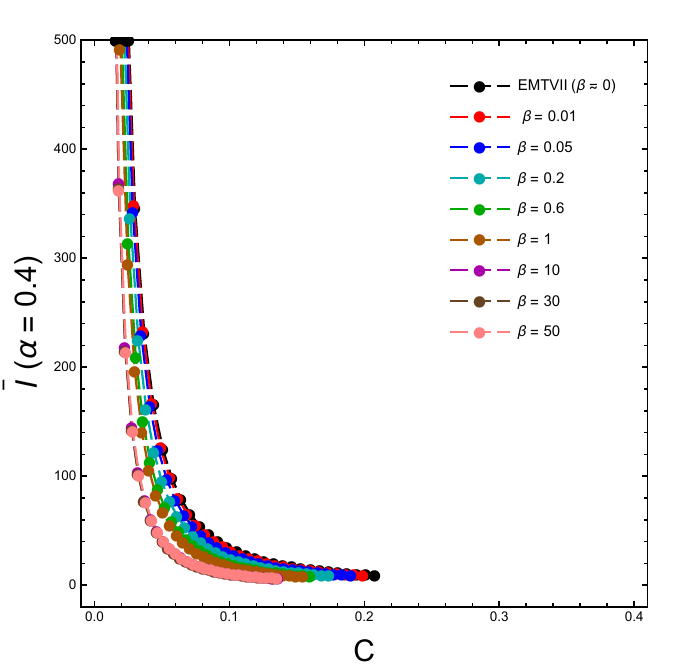}
	\includegraphics[width=1\linewidth]{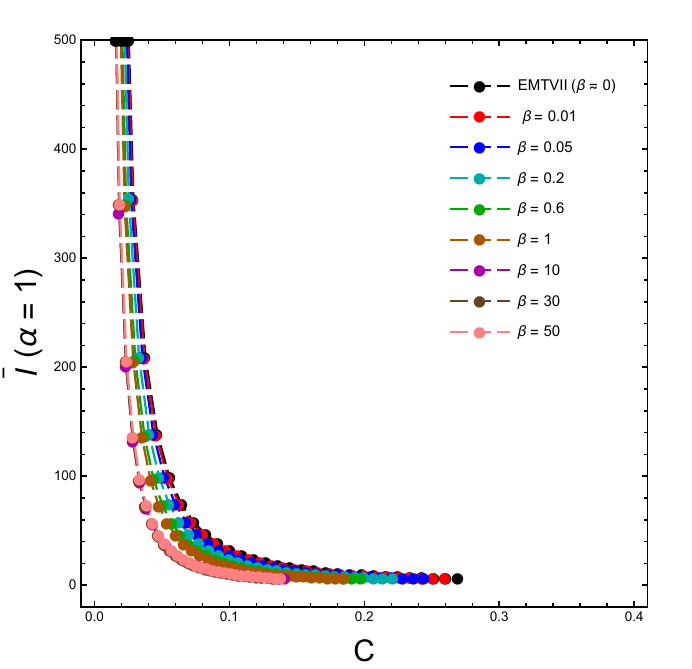}
	\caption{Dimensionless moment of inertia, $ \bar{I} $ vs $ \mathcal{C} $ for compactness within the causal limit. Each colour represents the profile for different value of $ \beta $ with black line (GR, $\beta=0$), red line ($ \beta=0.01 $), blue line ($ \beta=0.05 $), cyan line ($ \beta=0.2 $), dark orange line ($ \beta=1 $), magenta line ($ \beta=10 $), brown line ($ \beta=30 $), pink line ($ \beta=50 $)}
	\label{fig:9}
\end{figure}

\begin{figure}[t!]
	\centering
	\includegraphics[width=1\linewidth]{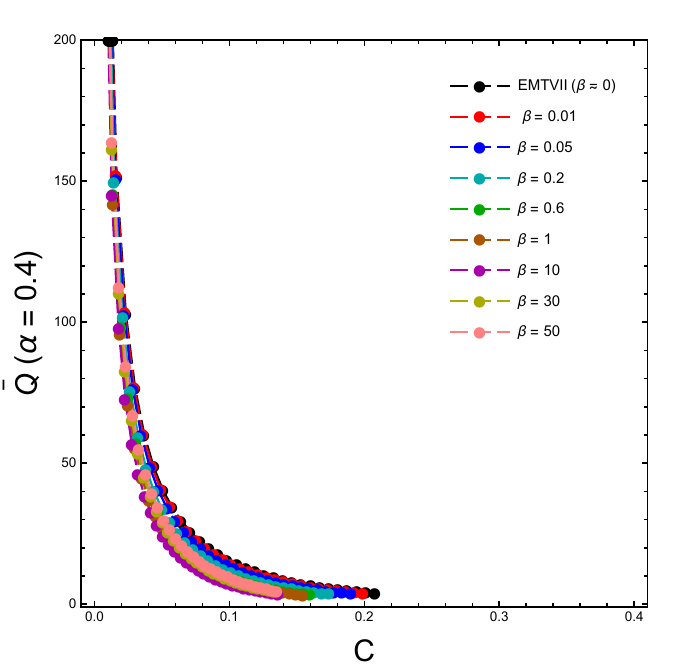}
	\includegraphics[width=1\linewidth]{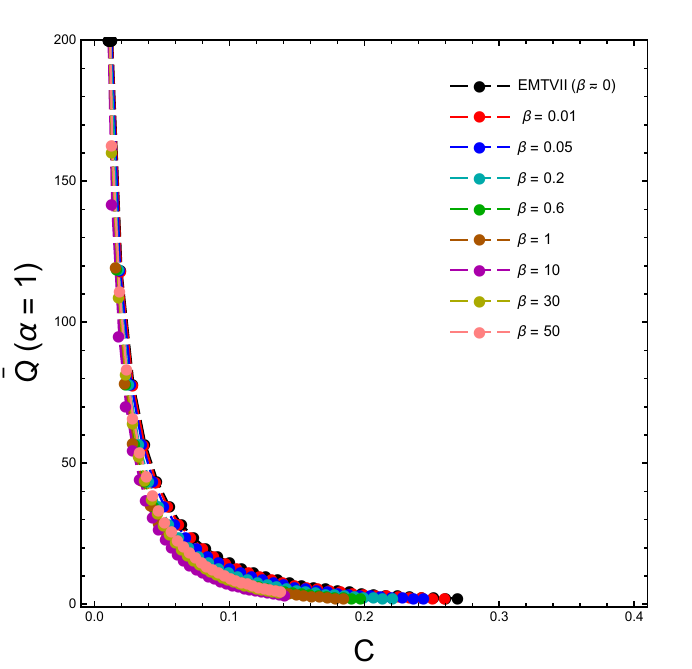}
	\caption{Dimensionless quadrupole moment, $ \bar{Q}, $ vs $ \mathcal{C} $ for the compactness within the causal limit. Each colour represents the profile for different value of $ \beta $ with black line (GR, $\beta=0$), red line ($ \beta=0.01 $), blue line ($ \beta=0.05 $), cyan line ($ \beta=0.2 $), dark orange line ($ \beta=1 $), magenta line ($ \beta=10 $), brown line ($ \beta=30 $), pink line ($ \beta=50 $)}
	\label{fig:111}
\end{figure}

\begin{figure}[t!]
	\centering
	\includegraphics[width=1\linewidth]{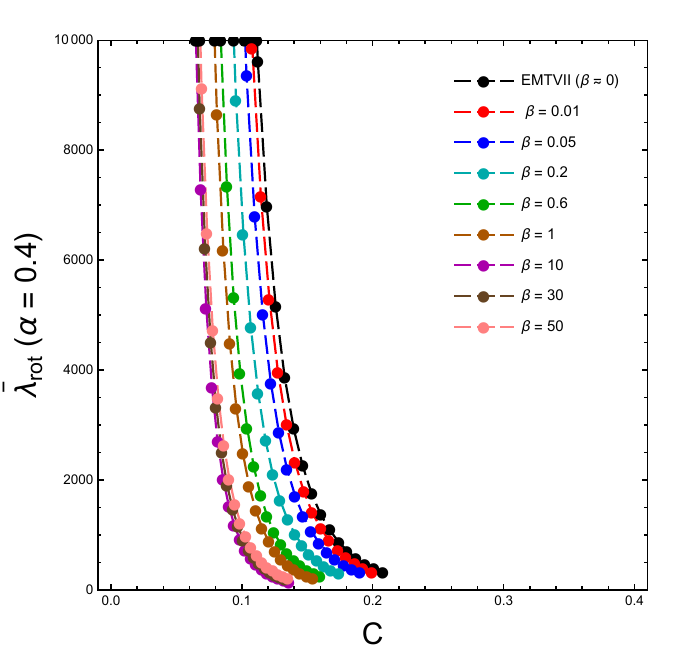}
	\includegraphics[width=1\linewidth]{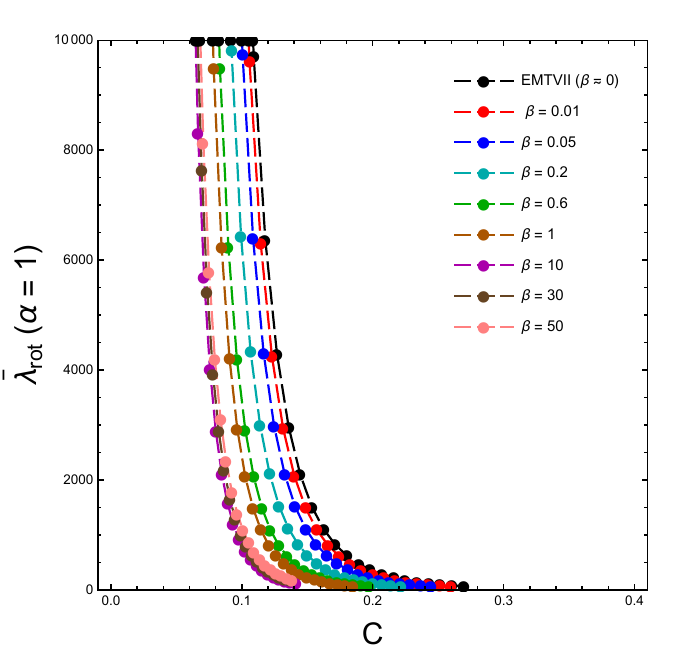}
	\caption{Dimensionless quadrupole moment, $ \bar{\lambda}_{\textrm{rot}} $ vs $ \mathcal{C} $ for the compactness within the causal limit. Each colour represents the profile for different value of $ \beta $ with black line (GR, $\beta=0$), red line ($ \beta=0.01 $), blue line ($ \beta=0.05 $), cyan line ($ \beta=0.2 $), dark orange line ($ \beta=1 $), magenta line ($ \beta=10 $), brown line ($ \beta=30 $), pink line ($ \beta=50 $)}
	\label{fig:13}
\end{figure}

We now turn our attention to the slowly rotating star profiles. 
At the linear order in the Hartle's expansion, the final equation \eqref{omegaeq} leads to the second-order differential equation for $\bar{\omega}$. The equation thus can be solved numerically by imposing the boundary condition at the center of the star. The characteristic of the profile can be seen from Figure \ref{fig:omega}. At the top part, we plot the radial profile of nonlocal angular velocity relative to the local inertial frame normalized by $\Omega$ with respect to $x$ for different parameters of the ratio between radius and mass of the star. We also plot the nonlocal angular velocity or the dragging inertial frame with the same normalized factor in the bottom part. In both figures, we vary the same parameter $x$ starting from the larger radius~(black color) and compare to the original Tolman VII case ($\alpha=1$)~\cite{Stuchlik:2021vdd,Posada:2023bnm}. Note that the solid color at the top figures is matched to that of the Tolman VII profile obtained in Ref.~\cite{Stuchlik:2021vdd}. For the $\bar{\omega}/\Omega$, we observe that each of the parameters has a minimum value at the origin and slowly goes up as $x$ increases. For the same inverse compactness, the nonlocal effect has a lower value than the GR case. This is a natural consequence of the higher central density in the nonlocal case. However, in the bottom figure, we have a flip profile since the expression reads $\frac{\omega}{\Omega} = 1-\frac{\bar{\omega}}{\Omega}$. It is shown that the angular velocity decreases from center to the surface. The frame dragging effect is strongest at the center and weakens outward. However, from this step, we can then continue to analyze our other profile by imposing the matching condition to get moment of inertia, $I=J/\Omega$. 

Since the results between finite pressure criteria and causal limit criteria show insignificant differences, i.e., the finite pressure constraint allows compact object with 20-30\% higher maximum mass. Here and henceforth, we will present only the results of the analyses under the causal limit which is more physical. Figure \ref{fig:9}-\ref{fig:13} show the relationships between the moment of inertia, quadrupole moment, rotational tidal deformability and compactness of the star. All quantities are decreasing functions of the compactness. As $ \beta $ increases, the maximum compactness decreases due to larger radius from the nonlocal smearing effect. At a given value of compactness, nonlocality reduces $\bar{I},\bar{Q},\bar{\lambda}_{\textrm{rot}}$ with respect to GR. However, the nonlocal effects are maximum around $\beta\simeq 10$ km$^2$.

\begin{figure}[t!]
	\centering
	\includegraphics[width=1\linewidth]{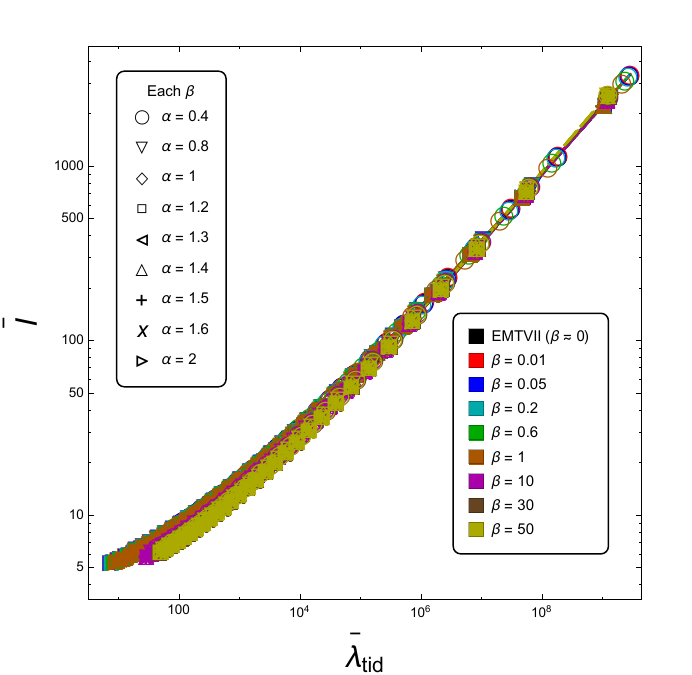}
	\includegraphics[width=1\linewidth]{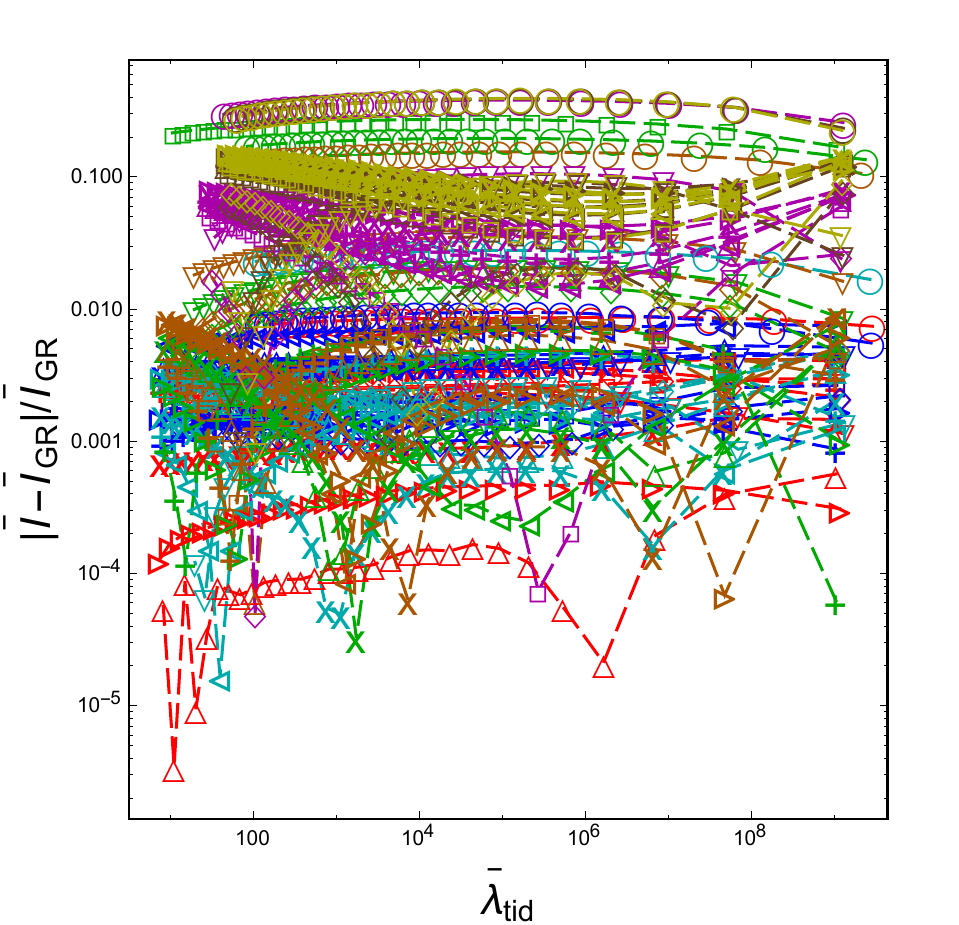}
	\caption{Dimensionless moment of inertia $ \bar{I} $ vs tidal $ \bar{\lambda} $ for [Top] $\mathcal{C}$ within causal limit. [Bottom] Fractional difference between the nonlocal and GR values. Plot for each value of $ \beta $ has varying values of $ \alpha $ designated with different shapes.}
	\label{fig:14b}
\end{figure}

\begin{figure}[t!]
	\centering
	\includegraphics[width=1\linewidth]{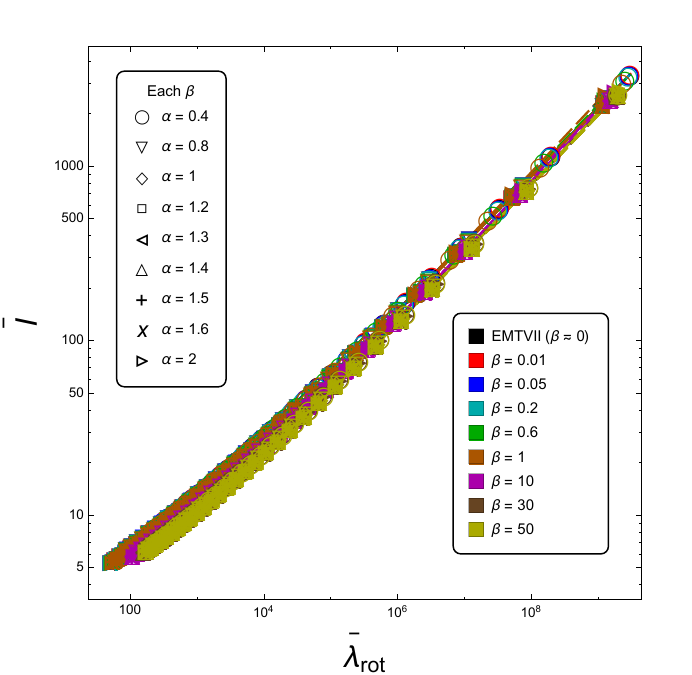}
	\includegraphics[width=1\linewidth]{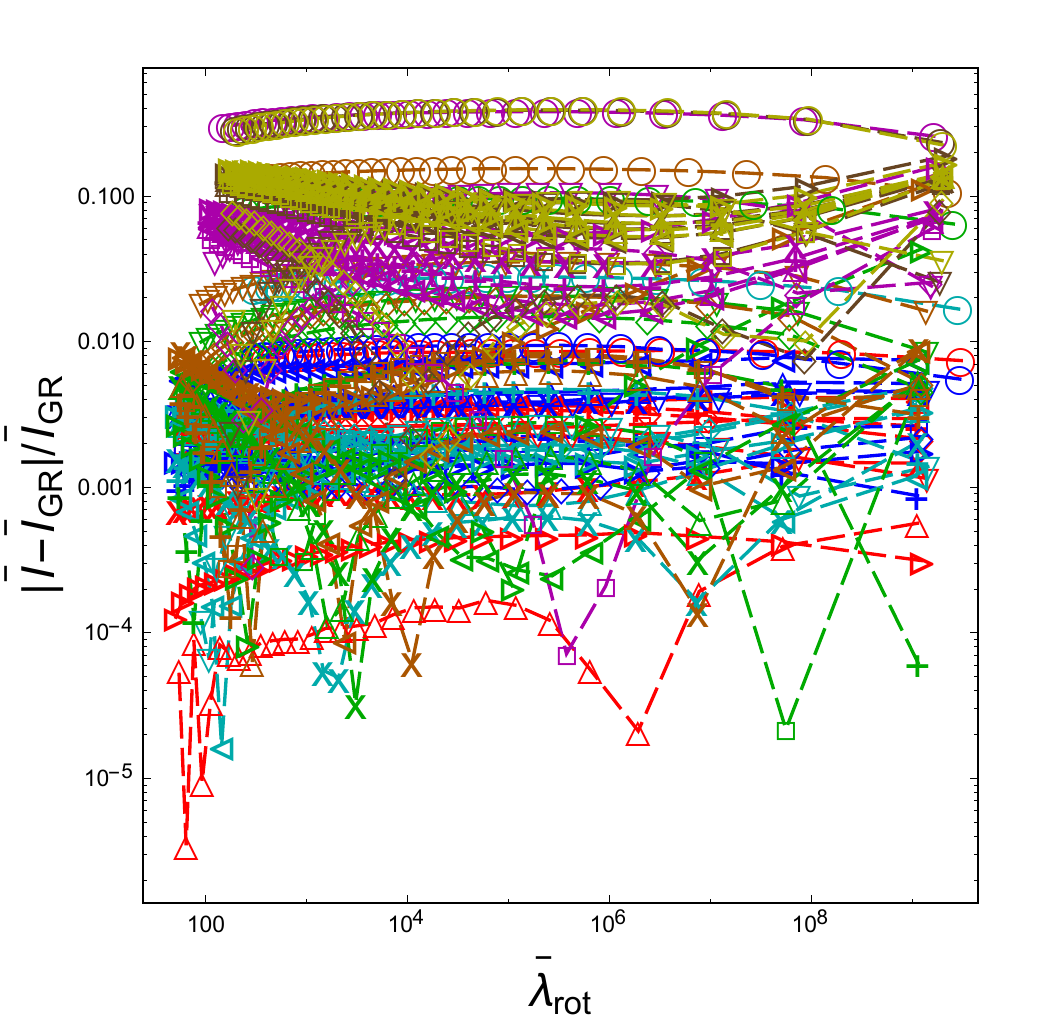}
	\caption{Dimensionless moment of inertia $ \bar{I} $ vs rotational tidal $ \bar{\lambda}_{\textrm{rot}} $ for [Top] $\mathcal{C}$ within causal limit. [Bottom] Fractional difference between the nonlocal and GR values. Plot for each value of $ \beta $ has varying values of $ \alpha $ designated with different shapes.}
	\label{fig:15b}
\end{figure}

\begin{figure}[t!]
	\centering
	\includegraphics[width=1\linewidth]{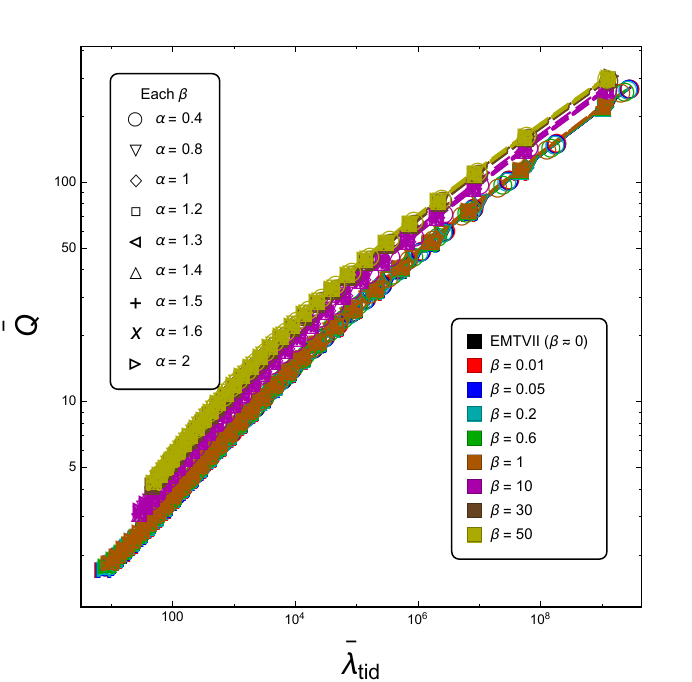}
	\includegraphics[width=1\linewidth]{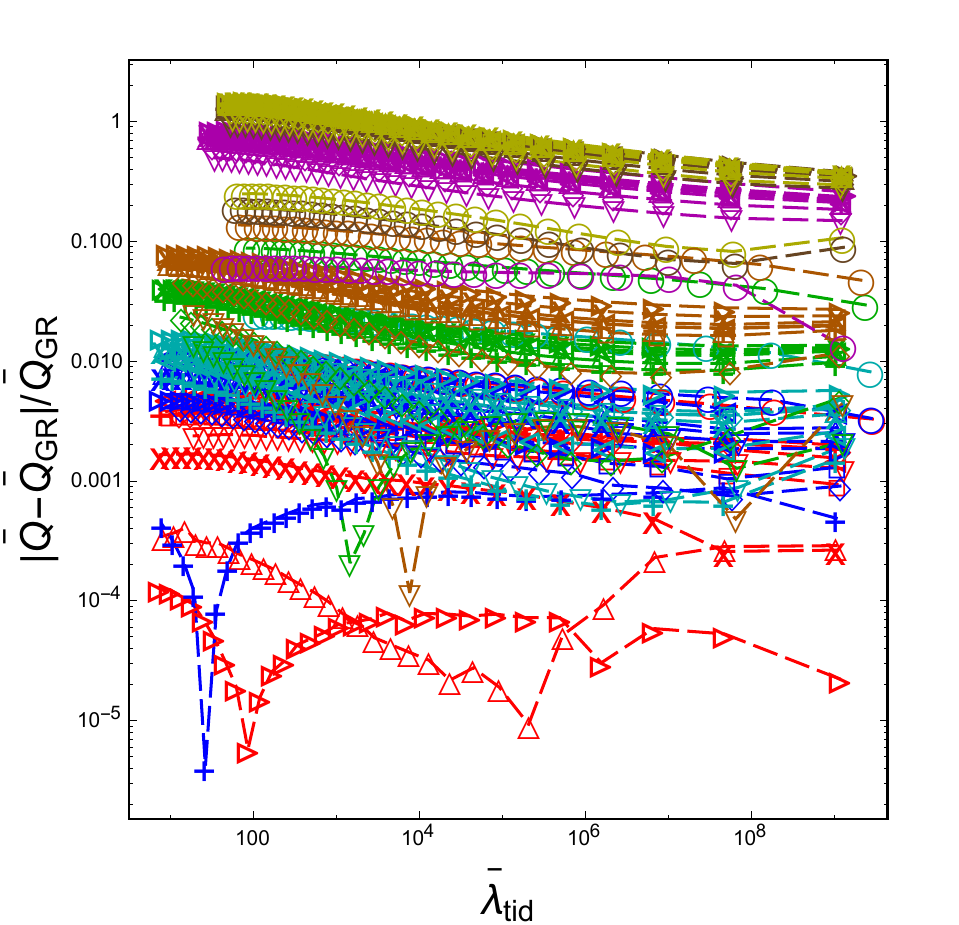}
	\caption{Dimensionless quadrupole deformation $ \bar{Q} $ vs tidal $ \bar{\lambda} $ for [Top] $\mathcal{C}$ within causal limit. [Bottom] Fractional difference between the nonlocal and GR values. Plot for each value of $ \beta $ has varying values of $ \alpha $ designated with different shapes.}
	\label{fig:16b}
\end{figure}

\begin{figure}[t!]
	\centering
	\includegraphics[width=1\linewidth]{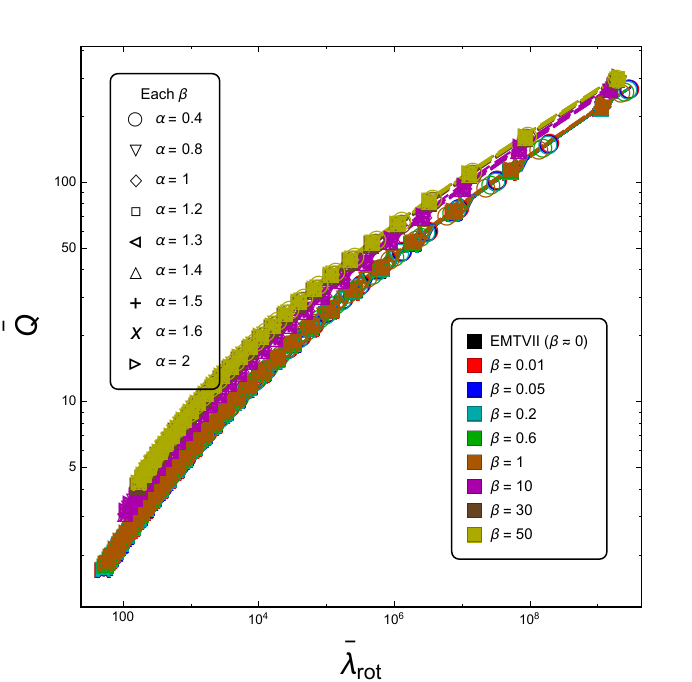}
	\includegraphics[width=1\linewidth]{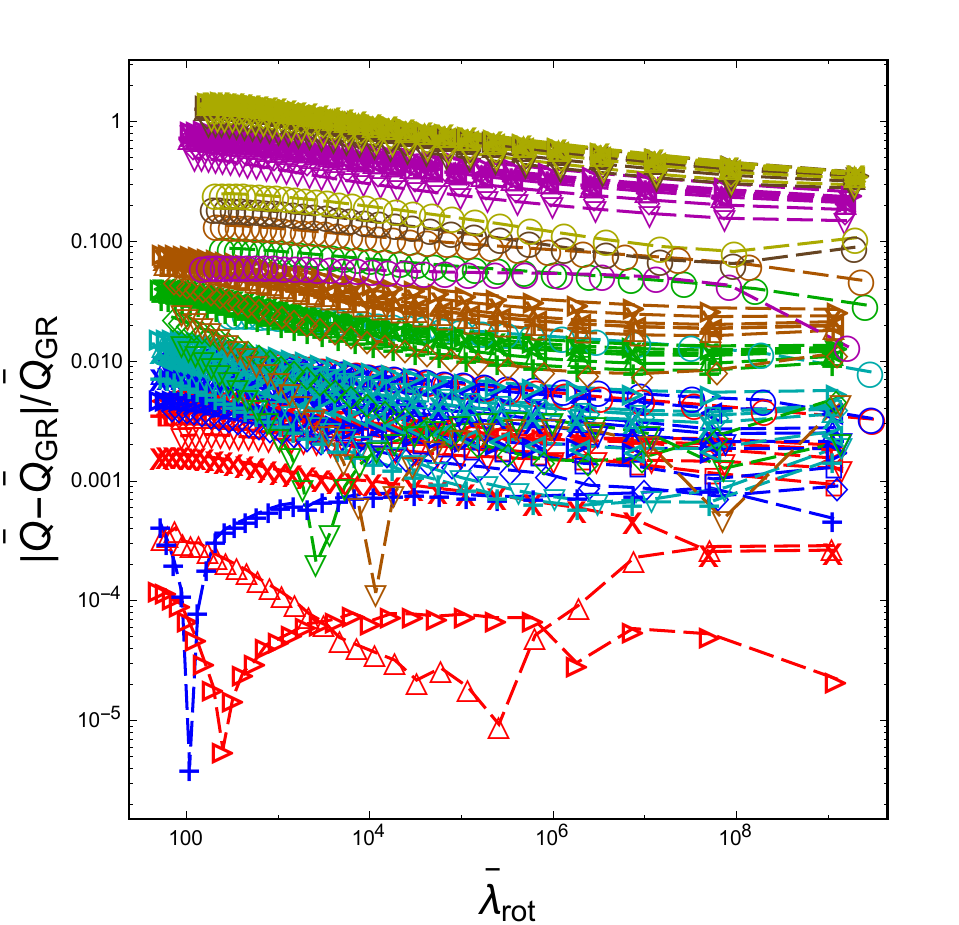}
	\caption{Dimensionless quadrupole deformation $ \bar{Q} $ vs rotational tidal $ \bar{\lambda}_{\textrm{rot}} $ for [Top] $\mathcal{C}$ within causal limit. [Bottom] Fractional difference between the nonlocal and GR values. Plot for each value of $ \beta $ has varying values of $ \alpha $ designated with different shapes.}
	\label{fig:17b}
\end{figure}

\begin{figure}[t!]
	\centering
	\includegraphics[width=1\linewidth]{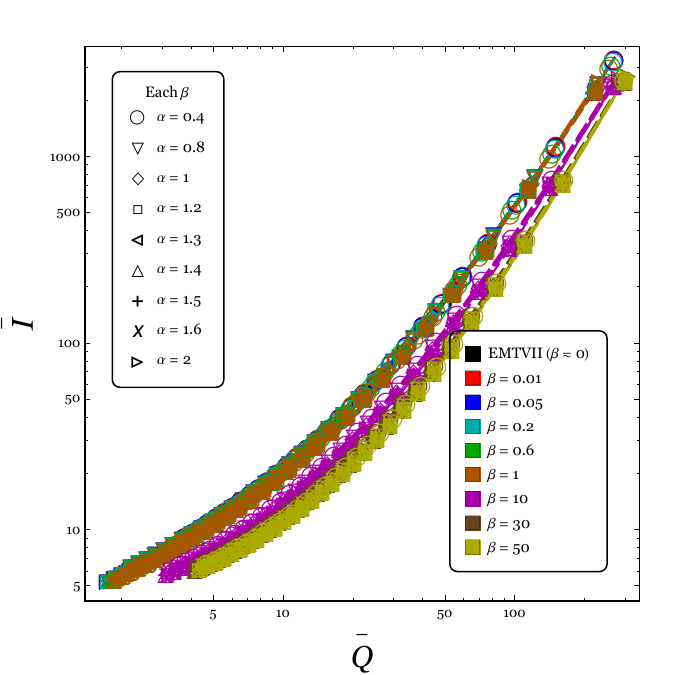}
	\includegraphics[width=1\linewidth]{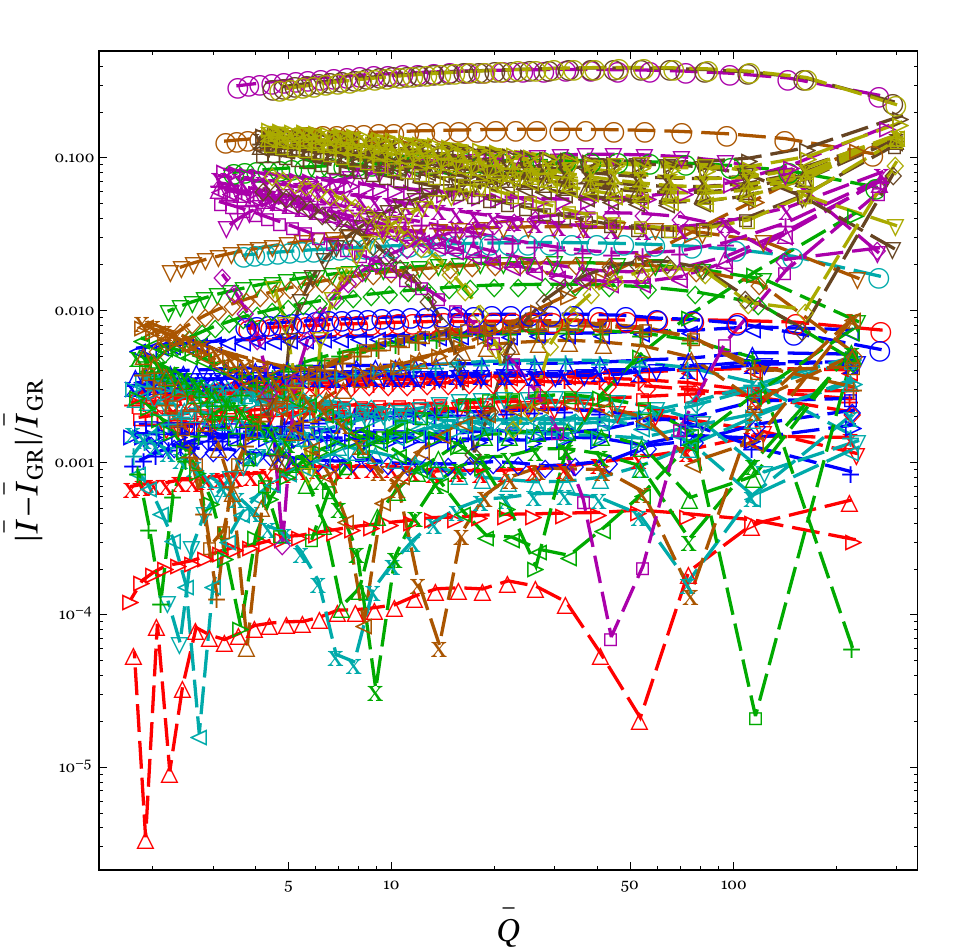}
	\caption{Dimensionless moment of inertia $ \bar{I} $ vs quadrupole deformation $ \bar{Q} $ for [Top] $\mathcal{C}$ within causal limit. [Bottom] Fractional difference between the nonlocal and GR values. Plot for each value of $ \beta $ has varying values of $ \alpha $ designated with different shapes.}
	\label{fig:18b}
\end{figure}

\begin{figure}[t!]
	\centering
	\includegraphics[width=1\linewidth]{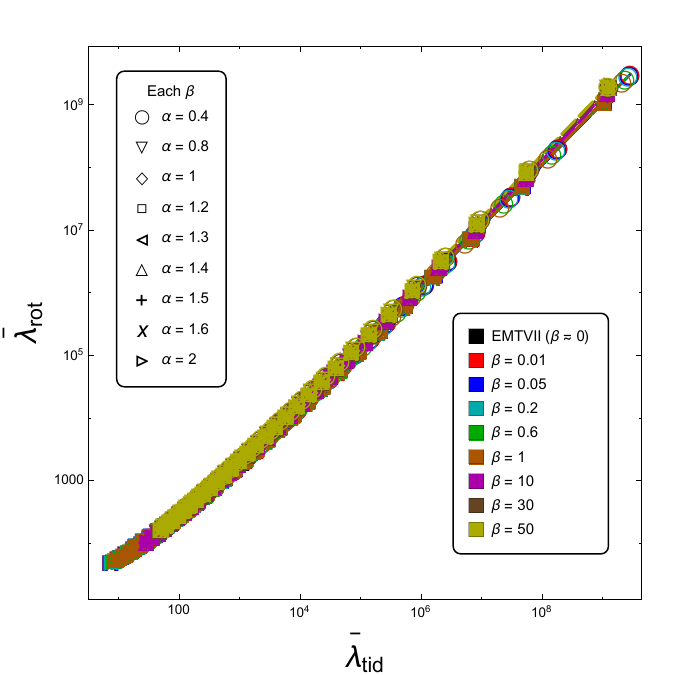}
	\includegraphics[width=1\linewidth]{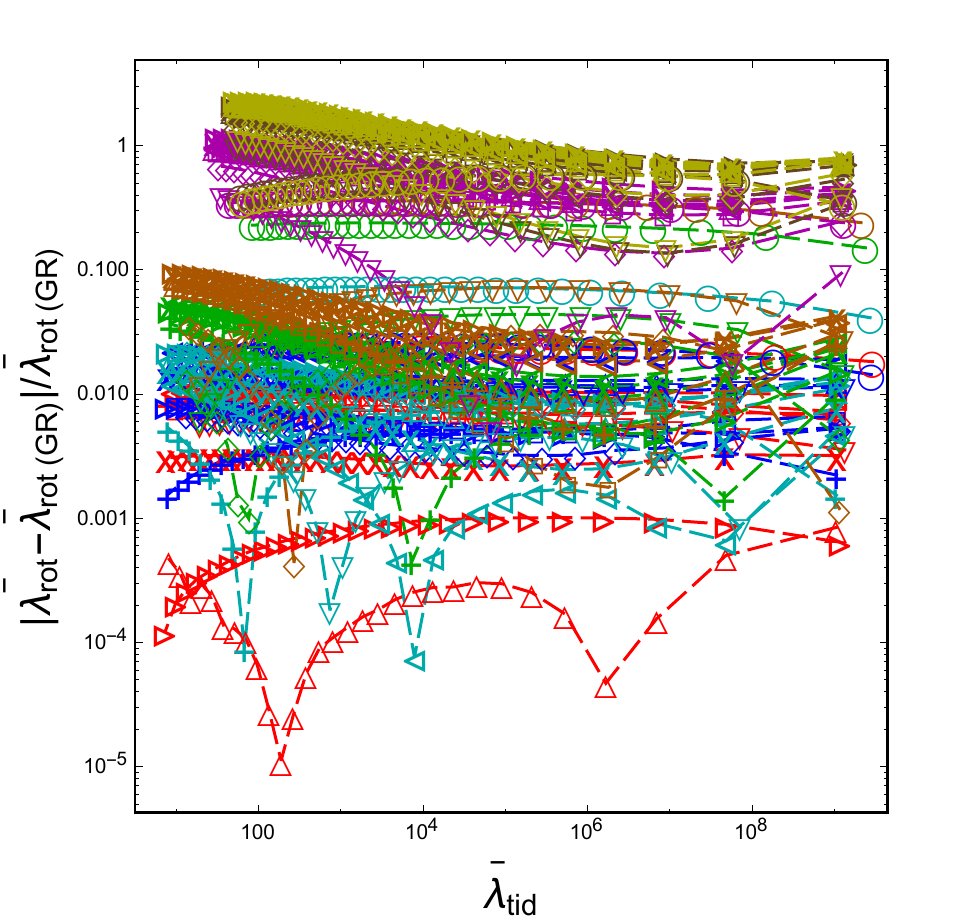}
	\caption{Dimensionless rotational tidal  $ \bar{\lambda}_{\textrm{rot}} $ vs tidal $ \bar{\lambda} $ for [Top] $\mathcal{C}\leq\mathcal{C}_{max}$ within causal limit. [Bottom] Fractional difference between the nonlocal and GR values. Plot for each value of $ \beta $ has varying values of $ \alpha $ designated with different shapes.}
	\label{fig:19b}
\end{figure}

Figures \ref{fig:14b}-\ref{fig:18b} illustrate the $I$-Love-$Q$ relation for various nonlocal parameters with the compactness based on the compactness within causal limit combined with the fractional difference plots. Plot for each value of $ \beta $ has varying values of $ \alpha $ designated with different shapes as shown in the figures. The fractional differences for each figure are calculated to show the deviation of the nonlocal model compared to GR values. Point to the left on the horizontal axis corresponds to the more massive star with higher compactness. Increasing nonlocal parameter makes the maximum compactness decreases. It can be seen from the shift profile in the plots. This is due to the influence of the nonlocal parameter on $ \tilde{\rho}_{\textrm{ext}} $ as shown in Eqn.~\eqref{rhoext}, which consequently increases the radius. On the other hand, from the figures, we can infer that the universality of these profiles are preserved for small values of nonlocal parameter $ \beta $, up to $ \beta = 1$ km$^2$. Interestingly, for each nonlocal parameter $ \beta $, the combination of free parameter $ \alpha $ shown in the figure are stacked in the same line representing the universality. On the other hand, we can also observe from each figures that the $I$-Love-$Q$ profiles approach the characteristic of black holes (BH) values as compactness increases (The values of $ \bar{I} $, $ \bar{\lambda}_{\textrm{tid}} $, $ \bar{\lambda}_{\textrm{rot}} $, and $ \bar{Q} $ for BH are $ \bar{I}\rightarrow 4 $, $ \bar{\lambda}_{\textrm{tid}} \rightarrow 0 $, $ \bar{\lambda}_{\textrm{rot}}\rightarrow 16 $, and $ \bar{Q}\rightarrow 1 $ \cite{Yagi:2013awa,Yagi:2013bca}). Nevertheless, the star profile with isotropic perfect fluid and the specified EoS cannot reach this BH limit ($ \mathcal{C}\rightarrow 0.5 $) by increasing the central density, or equivalenty the star compactness. Additionally, The Love-Love relations in Figure \ref{fig:19b} exhibit distinct profile when compared to the results presented in the other plots. These relation are stacking in the same line even we shift the nonlocal effect. Thus, we can infer that the $ \bar{\lambda}_{\textrm{rot}} - \bar{\lambda}_{\textrm{tid}} $ relation is approximately linear and weakly dependent on the EoS across the range of compactness values. One can also show that both tidally and rotationally-induced star deformabilities are exactly the same in the Newtonian limit \cite{Yagi:2013awa,Yagi:2013bca}. However, the deviation from the universality occurs when $\beta \gtrsim 10$ km$^2$ as shown in the fractional difference plot. 

\section{Conclusions and Discussions}   \label{Sec6}

We have investigated the Love number, tidal deformability, and slowly rotating profiles using both NEMTVII and EMTVII model. These models are distinguished by the nonlocal parameter $ \beta $ which are crucial for determining the compactness of the star. When the nonlocal parameter $ \beta $ is set to zero, the model reduces to the EMTVII model as proposed in \cite{Posada:2022lij}. Numerical analysis of these models \cite{Jayawiguna:2023vvw} indicate that the nonlocal parameter influences the exterior density profile ($ \tilde{\rho}_{\textrm{ext}} $), leading to an increase in the star radius due to the smearing effects.

The profile of the Love number ($k_{2}$) for the NEMTVII model decreases monotonically and approaches small value as the compactness ($ \mathcal{C} $) reaches its maximum value. The small value in Love number indicates that the star is centrally condensed (concentrated at the center) and stiff. Additionally, we have examined the implications of these findings by comparing the tidal deformability ($\bar{\lambda}_{\rm tid}, \bar{\lambda}_{\rm rot}$) with observational constraints from GW170817, GW190425, PSR J0348+0432, and PSR J0740+6620. The results indicate that the NEMTVII model, particularly with $ \alpha \gtrsim 1.6,\beta<1$ km$^2$, has the potential to remain within the constraints established by the observations of GW170817 and GW190425 as shown in Fig.~\ref{fig:7}.

We complete our study by evaluating $\bar{I},~\bar{Q},\bar{\lambda}_{\textrm{tid}},\bar{\lambda}_{\textrm{rot}} $ in both linear and second order in spin approximations within NEMTVII and EMTVII model. The calculations show that the nonlocal parameter $\beta$ smears a star and increases its radius (make it less compact) which also leads to distinctive characteristics of the nonlocal star in $\bar{I},~\bar{Q},\bar{\lambda}_{\textrm{tid}},\bar{\lambda}_{\textrm{rot}}$. Interestingly, for $ \beta \sim \mathcal{O}(1)$ km$^2$ nonlocality appears not to violate the universal relation proposed by Yagi-Yunes \cite{Yagi:2013awa,Yagi:2013bca}. Deviation from universal $I$-Love-$Q$ relation appears at $\beta \gtrsim 10$ km$^2$. 

In the nonlocal gravity theory, additional length scale $\ell_{\beta}\equiv \sqrt{\beta}$ is introduced to regulate spacetime for the distances smaller than $\ell_{\beta}$. In the black hole case, nonlocal smearing effects eliminate the curvature singularity of the Schwarzschild metric, replacing it with de Sitter core of the size $\sim \left(\ell_{\beta}^{3}/M l^{2}_{\rm Pl}\right)^{1/2}$~\cite{Modesto:2010uh}, where $M$ is the black hole mass and $l_{\rm Pl}$ is the Planck length. Interestingly, the regulation of spacetime singularity can be achieved regardless of the size of the nonlocal smearing scale $\ell_{\beta}$. The smearing scale could be macroscopic and much remote from the Planck scale and thus could lead to observable macroscopic effects. For astrophysical compact object, our analyses reveal that the nonlocal scale in the range $\ell_{\beta}\lesssim \mathcal{O}(1)$ km is consistent with LIGO constraints for NEMTVII profile with $\alpha \gtrsim  1.6$, and universal $I$-Love-$Q$ relation is preserved. When the nonlocal scale increases to the order of $\ell_{\beta} \gtrsim 3$ km, smearing effects become distinctive from GR and could be detectable astrophysically. The most distinctive effects of nonlocality are shown in the plots of $k_{2},\lambda_{\rm tid, rot}$ versus $C$ in Figure~\ref{fig:2},\ref{fig:3},\ref{fig:4},\ref{fig:5}, and \ref{fig:13} where the zeroth-order effects on the compactness are dominant. Deviation from the universal $I$-Love-$Q$ relation is another possible signature from the first and second order perturbations.

\newpage
\section{Acknowledgement}
We are grateful to Kent Yagi for insightful discussions and for highlighting the issue concerning the fractional plot. B.N.J. is supported by the Second Century Fund (C2F) and Research Abroad Scholarship, Chulalongkorn University, Thailand. P.B. is supported in part by National Research Council of Thailand~(NRCT) and Chulalongkorn University under Grant N42A660500.

~~~~~~~~~~~~~~~~~~~~~~~~~~~~~~~~~~~~~~~~~~~~~~~~~~~~~~~~~~~~~~~~~~~~~~~~~~~~~~~~~~~~~~~~~~~~~~~~~~~~~~~~~~
~~~~~~~~~~~~~~~~~~~~~~~~~~~~~~~~~~~~~~~~~~~~~~~~~~~~~~~~~~~~~~~~~~~~~~~~~~~~~~~~~~~~~~~~~~~~~~~~~~~~~~~~~~


\end{document}